\documentclass[preprint,showpacs,preprintnumbers,amsmath,amssymb]{revtex4}%
\usepackage{graphicx}
\usepackage{dcolumn}
\usepackage{bm}
\usepackage{amsmath}
\usepackage{amsfonts}
\usepackage{amssymb}%
\providecommand{\U}[1]{\protect\rule{.1in}{.1in}}
\newcommand{\be}{\begin{equation}}
\newcommand{\en}{\end{equation}}
\newcommand{\bea}{\begin{eqnarray}}
\newcommand{\ena}{\end{eqnarray}}
\begin{document}
\title{
Warm intermediate  inflationary Universe model in the presence of
a  Generalized Chaplygin Gas}
\author{Ram\'on Herrera}
\email{ramon.herrera@ucv.cl}
\affiliation{Instituto de F\'{\i}sica, Pontificia Universidad Cat\'{o}lica de
Valpara\'{\i}so, Avenida Brasil 2950, Casilla 4059, Valpara\'{\i}so, Chile.}
\author{Nelson Videla}
\email{nelson.videla@ing.uchile.cl}
\affiliation{Departamento de F\'{\i}sica, FCFM, Universidad de Chile, Blanco Encalada 2008, Santiago, Chile}
\author{Marco Olivares}
\email{marco.olivaresr@mail.udp.cl}
\affiliation{Facultad de Ingenier\'ia, Universidad Diego Portales,
Avenida Ej\'ercito Libertador 441, Casilla 298-V, Santiago, Chile.}
\date{\today}

\begin{abstract}
A warm intermediate inflationary model in the context of Generalized Chaplygin Gas is
investigated. We study this model in the weak
and strong dissipative regimes, considering a generalized form of the dissipative
coefficient $\Gamma=\Gamma(T,\phi)$, and we describe the
inflationary dynamics in the slow-roll approximation.
We find constraints on the  parameters in our model considering the Planck 2015
data, together with the condition for warm inflation $T>H$, and
the conditions for the weak and strong dissipative regimes.

\end{abstract}

\maketitle

%\preprint{GACG/07/2006}

%PACS, the Physics and Astronomy
%Classification Scheme.

\section{Introduction}

It is well known that in modern cosmology our understanding of the
early Universe has introduced  a new stage of the Universe,
called  the inflationary scenario
\cite{R1,R102,R103,R104,R105,R106}. This early phase solves some
of the problems of the standard big bang model, like the flatness,
horizon, density of monopoles, etc. However, the most important
feature of the inflationary scenario is that provides a
novel mechanism to account the large-scale structure
\cite{R2,R202,R203,R204,R205,Planck2015} and also explains the origin of the observed
anisotropy of the Cosmic Microwave Background (CMB)
radiation\cite{astro,astro2,astro202}.

On the other hand, in the warm inflation scenario,
the radiation production takes place at the same time that
inflationary expansion\cite{warm}. In this form, the presence of
radiation during the inflationary expansion implies that inflation
could smoothly end into the radiation domination epoch, without
introduce a reheating phase. In this way, the warm inflation
scenario avoids  the graceful exit problem. In the warm inflation
scenario, the dissipative effects are crucial during the
inflationary expansion, and these effects arise from a friction
term which drives the process of the scalar field dissipating into
a thermal bath. Originally  the idea of consider particle
production in the inflationary scenario was developed in
Ref.\cite{I1}, from the introduction of an anomalous dissipation
term in the equation of motion of the scalar field. However, the
introduction of the $\Gamma\dot{\phi}^2$ friction term in the
dynamics of the inflaton field $\phi$, as a source of radiation
production, was introduced in Ref.\cite{I2}, where $\Gamma$
corresponds to the dissipative coefficient. In fact, if the
radiation field is in an extremely excited state during the
inflationary epoch, and if there is a strong damping effect on the
inflaton dynamics, then it is obtained a strong dissipative
regime, and the otherwise is called the weak dissipative regime.

By the other hand, a fundamental condition for warm inflation to
occur is that the temperature of the thermal bath must satisfy
$T>H$, where $H$ is the Hubble rate. Under this condition, the
thermal fluctuations play a fundamental role in producing the
primordial density fluctuations, indispensable for large-scale
structure formation. In this sense, the thermal fluctuations of
the inflaton field predominates over the quantum ones
\cite{62526,1126}. For a review of warm inflation, see Ref.
\cite{Berera:2008ar}.

%%%%%%%%%%%%%%%%%%%%%%%%%%%%%%%%%%%%%%%%%%%%%%%%%%%%%%%%%%5
Also, it is well known that the Generalized Chaplygin Gas (GCG) is
other model that explains the  acceleration phase of the Universe.
The GCG has an exotic equation of state $p=p(\rho)$, given by
\cite{Chap}
\begin{equation}
p_{Ch}=-\frac{A}{\rho_{Ch}^{\beta}},  \label{r1}
\end{equation}
where $\rho_{Ch}$ and $p_{Ch}$ correspond to the energy density
and pressure of the GCG, respectively, and the quantities
%$0\leq\beta\leq 1$
$\beta$ and $A$ are constants. For the special case in which
$\beta$ = 1, this equation of state corresponds to the original
Chaplygin Gas \cite{Chap}, and the case of $\beta=0$ corresponds
to the $\Lambda$CDM model. From the perturbative analysis
considering the fluid version of the  GCG, negative values for
$\beta$ are not allowed, since the square of the speed of sound
$c_s^2=\beta\,A\rho^{-(\beta+1)} $, becomes negative, and
therefore this representation presents strong instabilities.
However, in the representation  of the GCG as a canonical
self-interacting scalar field, (where $c_s^2=1$) the perturbative
analysis can be performed even for negative values of
$\beta$\cite{F2}. Moreover, from the Supernova SN Ia analysis,
negative values for $\beta$ are favored when  the GCG is
considered  as a fluid\cite{F1}. In this form,  different
representations of the Chaplygin gas namely: as a fluid, tachyonic
field, a self-interacting scalar field or variant of gravity among
others, modify the constraints on the cosmological parameters, in
particular on the value of $\beta$. In the following, we will
consider any value of $\beta$, except the value $\beta=-1$,
since our  physical quantities and solutions present divergences.

 Considering the stress-energy
conservation equation and the Eq.(\ref{r1}), the energy density
can be written as
\begin{equation}
\rho_{Ch}=\left[A+\frac{B}{a^{3(1+\beta)}}\right]
^{\frac{1}{1+\beta}}=\rho_{Ch0}\left[A_s+\frac{(1-A_s)}{a^{3(1+\beta)}}\right]^
{\frac{1}{1+\beta}},\mbox{where}\,\,\,A_s=A/\rho_{Ch0}^{1+\beta}.
\label{r2}
\end{equation}
Here, $a=a(t)$ is the scale factor and the quantity $B$ is a
positive integration constant.
%
% {\bf From the solution given by
%Eq.(\ref{r2}), we  observe that this model behaves as a dust
%dominated universe at early times  and a De Sitter one at late
%times, and in between these two times, an intermediate epoch
%containing a mixture of vacuum energy density and a soft matter
%equation of state, $p=\beta \rho$, is present. As it was discussed
%in \cite{?}, the square of the speed of sound (fluid version)
%should not be greater than the speed of light during the soft
%matter stage. In this way, the perturbative analysis of this model
%haves a physical meaningful only if the allowed range for $\beta$
%is given by $0\leq\beta\leq1$.
%
%
% }
From the solution given by Eq.(\ref{r2}), the energy density of
the  GCG is characterized by two parameters, $A_s$ (or
equivalently $A$)  and $\beta$. The parameters $A_s$  and $\beta$
have been constrained from the observational data. In particular,
$A_s=0.73_{-0.06}^{+0.06}$ and $\beta=-0.09_{-0.12}^{+0.15}$ have
been obtained in Ref.\cite{const1}, the values $0.81\lesssim
A_s\lesssim 0.85$ and $0.2\lesssim\beta\lesssim 0.6$, have been
obtained in Ref.\cite{271}, and the constraints
$A_s=0.775_{-0.0161-0.0338}^{+0.0161+0.037}$,
$\beta=0.00126_{-0.00126-0.00126}^{+0.000970+0.00268}$, have been
obtained from the Markov Chain Monte Carlo method \cite{const2},
see also Ref.\cite{Avelino:2015dwa}.

In the construction of inflationary models inspired in the
Chaplygin Gas, the Eq.(\ref{r2}) can be extrapolate in the
Friedmann equation to study   an inflationary scenario
\cite{Bertolami:2006zg}. In this extrapolation, we identify the
energy density of matter with the contribution of the energy
density associated to the standard or tachyonic scalar field
\cite{271,mon1}. Specifically this modification is realized
from an extrapolation of Eq.(\ref{r2}), so that;
$\rho_{Ch}=\left[A+\rho_m^{(1+\beta)}\right]
^{\frac{1}{1+\beta}}\rightarrow
[A+\rho_\phi^{(1+\beta)}]^{\frac{1}{(1+\beta)}}$, where $\rho_m$
corresponds to the matter energy density and $\rho_\phi$
corresponds to the scalar field  energy density
\cite{Bertolami:2006zg}. In this form, the effective Friedmann
equation from the GCG may be viewed as a variant of gravity, which
presents a great interest in the study of the early Universe motivated
by string/M-theory\cite{mod}. In this
context, and in particular, if the effective Friedmann equation is
different to the standard Friedmann equation, then we consider it
to be a
 modified  gravity. In general  if the  field
equations   are  anything other than  Einstein's equations, or
action, then we view it to be a modified theory of gravity.  For a
review of modified gravity theories and cosmology, see e.g., Ref.\cite{mod2}.

In the context of exact solutions, an expansion of the power-law type can be
found from  an exponential potential, where the scale factor
evolves as $a(t)\sim t^{p}$, where
$p>1$\cite{power}. de Sitter inflation is other exact solution to
the background equations, which can be obtained from a constant
effective potential \cite{R1}.
 However, another type of  exact
solution corresponds to intermediate
inflation, where the expansion rate
is slower than de Sitter inflation, but faster than power-law
inflation. In this model, the scale factor $a(t)$ evolves as
\begin{equation}
a(t)=\exp[\,\alpha\,t^{f}],  \label{at}
\end{equation}
where $\alpha$ and $f$ are two constants; $\alpha>0$ and $0<f<1$
\cite{Barrow1}.

The model of intermediate inflation was in the beginning formulated as an exact
solution to the background equations, nevertheless this model  may
be studied under the slow-roll approximation together with
the cosmological perturbations. In particular, under the slow-roll
analysis, the effective potential is a power law type, and the scalar spectral index becomes $n_s\sim 1$, and exactly
$n_s=1$ (Harrizon-Zel'dovich spectrum) for the special value
$f=2/3$
 \cite{Barrow2}. In the same way, the tensor-to-scalar
ratio $r$ becomes $r\neq 0$\cite{ratior,Barrow3}. Also, other
motivation to study intermediate inflation comes from
string/M-theory{\cite{KM,ART}} (see also,
Refs.\cite{BD,Varios1,Varios2,Sanyal}). Here, is possible  to resolve
the initial singularity and also to give account of the present
acceleration of the universe, among others \cite{new1,varios1}.

%%%%%%%%%%%%%%%%%%%%%%%%%%%%%%%%%%%%%%%%55

%%%%%%%%%%%%%%%%%%%%%%%%%%%%%%%%%%%%%%%%%%%%%%%

The main goal of the present work is to study the development of
an intermediate-GCG model in the context of warm inflation. To
achieve this, we will not view the solution given by Eq.(\ref{r2})
as a  result of the adiabatic fluid from Eq.(\ref{r1}), and
therefore as a fluid representation, but rather,  recognizing the
energy density of matter  as the contribution  of the energy
density associated with a standard  scalar field.  From this
perspective
 we will obtain a modified Friedmann equation, and we will analyze the GCG
as a representation from of the variant of gravity \cite{Bertolami:2006zg}.
From this modification itself, we will study the warm inflation scenario, and
we will consider that this model presents dissipative effects coming from an
interaction between a standard scalar field and a radiation field. In relation to the friction term,
 we consider a generalized form of the dissipative
coefficient $\Gamma=\Gamma(T,\phi)$, and we study how it
influences the inflationary dynamics. In this form, we will study
the background dynamics and the cosmological perturbations for our
model in two regimes, namely the weak and strong dissipative
regimes. Also, we find constraints on parameters of our model
considering the new data of Planck 2015 \cite{Planck2015},
together with the condition for warm inflation, given by $T>H$,
and the conditions for weak ($\Gamma<3H$) and strong ($\Gamma>3H$)
dissipative regimes.

%%%%%%%%%%%%%%%aca

The outline of the paper is as follows: The next section presents
a short description of  the warm intermediate inflationary model
in the context of the GCG. In the sections III and VI, we discuss the
warm-GCG model in the weak and strong dissipative regimes. In each
section, we find explicit expressions for the dissipative
coefficient, scalar potential, scalar power spectrum and
tensor-scalar ratio. Finally, section V resumes our finding and exhibits our
conclusions. We chose units so that $c=\hbar=1$.

\section{The Warm Inflationary phase and the GCG.\label{secti}}

During warm inflation, the Universe is
filled with a self-interacting scalar field with energy density
$\rho_{\phi}$ together with a radiation field of energy density
$\rho_{\gamma}$. In this way, the total energy density
$\rho_{total}$ corresponds to
$\rho_{total}=\rho_\phi+\rho_\gamma$. In the following, we will
consider that the energy density $\rho_{\phi}$ associated to the
scalar field  is defined as
$\rho_{\phi}=\dot{\phi}^{2}/2+V(\phi)$ and the pressure as
$P_{\phi}=\dot{\phi}^{2}/2-V(\phi)$, where $V(\phi)$ corresponds
to the effective scalar potential.

On the other hand, the GCG model can be also considered to
achieve an inflationary scenario from  the modified Friedmann
equation, given by \cite{271}
\begin{equation}
H^{2}={\kappa \over 3}
\left(\left[A+\rho_{\phi}^{1+\beta}\right]^{1 \over
1+\beta}+\rho_{\gamma}\right).
\label{HC}%
\end{equation}
Here $H$ corresponds to the
Hubble rate, defined as   $H=\dot{a}/a$, and the constant $\kappa=8\pi G=8\pi/m_{p}^{2}$ ($m_{p}$ denotes
the Planck mass). Dots mean derivatives with respect to cosmic time.

The modification in the Friedmann equation given by Eq.(\ref{HC}),
is the so-called Chaplygin-inflation\cite{271}.  In this form,
 the GCG inflationary model may be viewed as a modification of the gravity according to
 Eq.(\ref{HC}).

 The dynamical equations for the energy densities
$\rho_{\phi}$ and $\rho_{\gamma}$ in the warm inflation
scenario are given  by\cite{warm}
\begin{equation}
\dot{\rho_{\phi}}+3\,H\,(\rho_{\phi}+P_{\phi})=-\Gamma\;\;\dot{\phi}^{2},\,\,\,\,\mbox{or
equivalenty}\,\,\,\ddot{\phi}+3H\dot{\phi}+V'=-\Gamma\dot{\phi} ,
\label{key_01}%
\end{equation}
and
\begin{equation}
\dot{\rho}_{\gamma}+4H\rho_{\gamma}=\Gamma\dot{\phi}^{2}, \label{key_02}%
\end{equation}
where, $V'=\partial V/\partial\phi$ and $\Gamma>0$ corresponds to
the dissipative coefficient. It is well known that the coefficient
$\Gamma$, is responsible of the decay of the scalar field into
radiation. In general,  this coefficient can be
assumed  to be a constant or a function of the temperature $T$ of
the thermal bath $\Gamma(T)$, or the scalar field $\phi$, i.e.,
$\Gamma(\phi)$, or also both $\Gamma(T,\phi)$\cite{warm}. A
general form for the dissipative coefficient $\Gamma(T,\phi)$ is
given by\cite{gamma1}

\begin{equation}
\Gamma(T,\phi)=C_{\phi}\,\frac{T^{m}}{\phi^{m-1}}, \label{G}%
\end{equation}
where  the constant  $C_\phi$ is associated with  the
microscopic dissipative dynamics, and the value $m$ is an integer. Depending of
the different values of $m$,  the dissipative coefficient given by
Eq.(\ref{G}) includes  different cases
\cite{gamma1}. In particular, for the
value of $m=3$, or equivalently $\Gamma= C_\phi T^3\phi^{-2}$, has
been studied in Refs.\cite{new2,26,28,2802}. For the
cases $m=1$, $m=0$ and $m=-1$, the dissipative coefficient is
related to supersymmetry and non-supersymmetry cases\cite{gamma1,28}.

Considering that  during the scenario of  warm inflation the
energy density associated to the scalar field $\rho_{\phi}\gg
\rho_{\gamma}$\cite{warm,62526,6252602,6252603,6252604}, i.e., the
energy density of the scalar field predominates  over the energy
density of the radiation field, then  the Eq.(\ref{HC}) may be
written as
\begin{equation}
H^{2}\approx {\kappa \over
3}\,\left(A+\rho_{\phi}^{1+\beta}\right)^{ 1\over 1+\beta}
={\kappa \over 3}\,\left[A+\left({\dot{\phi}\over
2}+V(\phi)\right)^{1+\beta}\right] ^{ 1\over 1+\beta}.\label{inf2}
\end{equation}

Now, combining Eqs. (\ref{key_01}) and (\ref{inf2}), the quantity
$\dot{\phi}^2$ becomes
\begin{equation}
\dot{\phi}^{2}= {\frac{2 }{\kappa}}\frac{(-\dot{H})}{(1+R)}\,
\left[1-A\left({3\,H^2 \over \kappa}\right)^{-(1+\beta)}\right]^{-
\beta\over 1+\beta}
,\label{inf3}%
\end{equation}
where the parameter $R$ corresponds to the ratio between $\Gamma$
and the Hubble rate, which is defined as
\begin{equation}
R=\frac{\Gamma}{3H}, \label{rG}%
\end{equation}
we note that for the case of the weak dissipative regime, the
parameter  $R<1$ i.e., $\Gamma<3H$, and during the  strong
dissipation regime, we have  $R>1$ or equivalently $\Gamma>3H$.

We also consider  that the radiation production is quasi-stable, then   $\dot{\rho
}_{\gamma}\ll4H\rho_{\gamma}$ and $\dot{\rho}_{\gamma}\ll\Gamma\dot{\phi}^{2}%
$, see Refs.\cite{warm,62526,6252602,6252603,6252604}.  In this
form, combing Eqs.(\ref{key_02}) and (\ref{inf3}), the energy density for the radiation
field can be written as
\begin{equation}
\rho_{\gamma}=\frac{\Gamma\dot{\phi}^{2}}{4H}=\frac{\Gamma(-\dot{H})}{2\kappa
H(1+R)} \,\left[1-A\left({3\,H^2 \over
\kappa}\right)^{-(1+\beta)}\right]^{- \beta\over
1+\beta}=C_{\gamma}\,T^{4},
\label{rh}%
\end{equation}
where the quantity  $C_{\gamma}%
=\pi^{2}\,g_{\ast}/30$, in which $g_{\ast}$ denotes the number of
relativistic degrees of freedom. In particular, for the Minimal Supersymmetric Standard Model
(MSSM), $g_{\ast}= 228.75$ and $C_{\gamma} \simeq 70$ \cite{62526}.

From Eq.(\ref{rh}), we get that the temperature of the thermal
bath $T$, is given by
\begin{equation}
T=\left[
\frac{\Gamma\,(-\dot{H})}{2\,\kappa\,\,C_{\gamma}H\,(1+R)}\right]^{1/4}
\,\left[1-A\left({3\,H^2 \over
\kappa}\right)^{-(1+\beta)}\right]^{- \beta\over 4(1+\beta)},
\label{rh-1}%
\end{equation}
and considering  Eqs.(\ref{inf2}), (\ref{inf3}) and (\ref{rh}) the
effective  potential becomes

\begin{equation}
V=\left[\left({3\,H^2 \over
\kappa}\right)^{1+\beta}-A\right]^{1\over 1+\beta}
+\frac{\dot{H}(2+3R)}{2 \kappa(1+R)} \,\left[1-A\left({3\,H^2
\over \kappa}\right)^{-(1+\beta)}\right]^{- \beta\over 1+\beta}.
\label{pot}%
\end{equation}
Here, we note that this effective potential could be expressed in
terms of the scalar field, in the case of the weak (or strong)
dissipative regime.

Similarly, combining  Eqs.(\ref{G}) and (\ref{rh-1}) the
dissipation coefficient $\Gamma$, may be written as
\begin{equation}
\Gamma^{{\frac{4-m }{4}}}=\,C_{\phi}\,\phi^{1-m} \left[
\frac{-\dot{H}}{2\kappa\, C_{\gamma}H(1+R)}\right]  ^{m/4}\,
\left[1-A\left({3\,H^2 \over \kappa}\right)^{-(1+\beta)}\right]
^{-m\, \beta\over 4(1+\beta)}. \label{G1}%
\end{equation}
 Here, the
Eq.(\ref{G1}) determines the dissipative coefficient in the weak
(or strong) dissipative regime in terms of the scalar field (or
the cosmic time).

%\section{ Intermediate inflation.\label{section3}}

In the following, we will analyze our warm Generalized Chaplygin Gas model in the
context of intermediate inflation. To achieve this, we will consider  a
general form of the dissipative coefficient $\Gamma$ given by
Eq.(\ref{G}), for the specific cases $m = 3, m = 1. m = 0$ and $m
= -1$. Also, we will restrict ourselves to the weak and  strong
dissipative regimes.

\section{ The weak dissipative regime.\label{subsection1}}

We start by considering that our model evolves
according to the weak dissipative regime, in which  $\Gamma<3H$.
In this way, the standard scalar field $\phi $ as function of cosmic time,
from Eqs.(\ref{at}) and (\ref{inf3}), is found to be
\begin{equation}
\phi (t)-\phi _{0}=\frac{B[t]}{K},  \label{fisol}
\end{equation}%
where $\phi(t=0)=\phi_0$ corresponds to an
integration constant, and $K$ is a constant given by
$$
K=(1+\beta)\, \sqrt{6\,(1-f)}\,\left({\kappa\over 3}\right)^{\frac{2-f}{4(1-f)}}\,
(\alpha\,f)^{\frac{-1}{2(1-f)}}\,A^{\frac{f}{4(1+\beta)(1-f)}}, %
$$
and  $B[t]$, denotes the incomplete Beta function
\cite{Libro}, defined as
$$
B[t]= B\left[A\,\left({\kappa\over
3\alpha^2f^2}\right)^{1+\beta}\,t^{2(1+\beta)(1-f)}
;\frac{f}{4(1+\beta)(1-f)},\frac{2+\beta}{2(1+\beta)}\right].
$$

In the following we will assume the integration constant
$\phi_{0}=0$ (without loss of generality). From the solution of
the scalar field given by Eq.(\ref{fisol}), the Hubble rate $H$ in
terms  of the scalar field becomes $
H(\phi)=\alpha\,f\,\left(B^{-1}[K\,\phi]\right)^{-(1-f)}$, where
$B^{-1}[K\,\phi]$ represents the inverse of the function $B[t]$.

Considering the slow-roll approximation in which
$\dot{\phi}^2/2<V(\phi)$, then from Eq.(\ref{pot}) the scalar
potential as function of the scalar field, can be written as
\begin{equation}
V(\phi)\approx \left[\left({3\,\alpha^2f^2\over
\kappa\left(B^{-1}[K\,\phi]\right)^{2(1-f)}}\right)^{1+\beta}
-A\right]^{1\over 1+\beta}.
\label{pot11}%
\end{equation}

Assuming that the model evolves according to the weak dissipative regime, then   the
dissipative coefficient $\Gamma$ as function of the scalar field,
for the case of $m \neq 4$,
 results
\begin{equation}
\Gamma(\phi)=\,C_{\phi}^{{\frac{4}{4-m}}}\left[  \frac{1-f}%
{2\kappa\,C_{\gamma}\,B^{-1}[K\,\phi]}\right]^{\frac{m}{4-m}}\phi^{{\frac{4(1-m)
}{4-m}}}
\,\left[1-A\left({\kappa\left(B^{-1}[K\,\phi]\right)^{2(1-f)}
\over3\,\alpha^2f^2}\right) ^{(1+\beta)}\right] ^{- m\,\beta\over
(4-m)(1+\beta)}
, \label{gammaph}%
\end{equation}
here, we have considered  Eq.(\ref{G1}).

On the other hand, we obtain that the dimensionless slow-roll
parameter $\varepsilon$, from Eq.(\ref{G1}) is given by $
\varepsilon=-\frac{\dot{H}}{H^{2}}=\left(  {\frac{1-f}{Af}}\right)
{1\over (B^{-1}[K\,\phi])^{f}}$. In this way, the condition
$\varepsilon<$1 (condition for inflation to occur) is satisfied
for values of the scalar field, such that; $\phi>{1\over
K}B\left[\left({\frac{1-f}{Af}}\right)^{1/f}\right]$.

%and the other slow-roll parameter  becomes
%$
%\eta=-\frac{\ddot{H}}{H\dot{H}}=k_{0}^{2}\left(  {\frac{2-f}{Af}}\right)
%\phi^{-2}\,. \label{eta}%
%$

%\bigskip
From the definition of  the number of $e$-folds $N$ between two
different values of cosmic time, $t_{1}$ and $t_{2}$, or
between two values of the scalar field, namel $\phi_{1}$ and
$\phi_{2}$, is given by
 \begin{equation}
N=\int_{t_{1}}^{t_{2}}\,H\,dt=\alpha\,\left(
t_{2}^{f}-t_{1}^{f}\right)
=\alpha\,\left[(B^{-1}[K\,\phi_{2}])^{f}-(B^{-1}[K\,\phi_{1}])^{f} \right]. \label{N1}%
\end{equation}
Here, we have used Eq.(\ref{fisol}).

The  inflationary scenario begins at the earliest stage possible,
for which  $\varepsilon=1$, see Ref.\cite{Barrow2}. In this form, from
the definition of the parameter $\varepsilon$, the value of the scalar field
$\phi_{1}$ results
\begin{equation}
\phi_{1}={1\over K}B\left[\left({\frac{1-f}{Af}}\right)^{1/f}\right]\;. \label{al}%
\end{equation}

In the following we will analyze the scalar and tensor
perturbations during the weak dissipative regime ($R<1$) for our
Chaplygin warm model. It is well known that
 the density perturbation may be written as ${\mathcal{P}%
_{\mathcal{R}}}^{1/2}=\frac{H}{\dot{\phi}}\,\delta\phi$\cite{warm}.
However, during the warm inflation scenario, a
thermalized radiation component is present, so the inflation
fluctuations are principally thermal instead quantum
\cite{warm,62526,6252602,6252603,6252604}. In fact,
for the weak dissipation regime, the inflaton fluctuation
$\delta\phi^{2}$ is found to be $\delta\phi^{2}\simeq
H\,T$ \cite{62526,6252602,6252603,6252604,B1}. Therefore, the
power spectrum of the scalar perturbation
${\mathcal{P}_{\mathcal{R}}}$, from Eqs.(\ref{inf3}), (\ref{rh-1})
and (\ref{G1}), becomes

\begin{equation}
{\mathcal{P}_{\mathcal{R}}}={\sqrt{3\pi}\kappa\over 4}\, \left(
\frac{C_{\phi}}{2\kappa C_{\gamma}}\right)
^{{\frac{1}{4-m}}}\phi^{{\frac{1-m}{4-m}}}H^{{\frac
{11-3m}{4-m}}}(-\dot{H})^{-{\frac{3-m}{4-m}}}
\left[1-A\left({3\,H^2 \over \kappa}\right)^{-(1+\beta)}\right]
^{ \beta(3-m)\over (1+\beta)(4-m)}, \label{pd}%
\end{equation}
or equivalently the power spectrum of the scalar perturbation may
be expressed in terms of the scalar
 field as
\begin{equation}
{\mathcal{P}_{\mathcal{R}}}=k_{1}\,\,\phi^{\,\frac{1-m}{4-m}}
\left(B^{-1}[K\,\phi] \right)  ^{\frac{2f(4-m)+m-5}{4-m}} \left[
1-A\left({\kappa\left(B^{-1}[K\,\phi]\right)^{2(1-f)}
\over3\,\alpha^2f^2}\right)
^{(1+\beta)} \right]^{ \beta(3-m)\over (1+\beta)(4-m)}, \label{pd}%
\end{equation}
where the constant $k_{1}$ is defined as
$
k_{1}={\sqrt{3\pi}\kappa\over 4}\left(  \frac{C_{\phi}}{2\kappa
C_{\gamma}}\right)
^{{\frac{1}{4-m}}}(\alpha\,f)^{2}%
\,(1-f)^{{\frac{m-3}{4-m}}} $.

Also, the power spectrum may be written as function of the number
of $e-$folds $N$, obtaining

\begin{equation}
{\mathcal{P}_{\mathcal{R}}}(N)=k_{2}\,\,(B[J(N)])^{\,\frac{1-m}{4-m}}
\left(J[N] \right)  ^{\frac{2f(4-m)+m-5}{4-m}} \left[
1-A\left({\kappa\left(J[N]\right)^{2(1-f)}
\over3\,\alpha^2f^2}\right) ^{(1+\beta)} \right]^{ \beta(3-m)\over
(1+\beta)(4-m)}.
\label{pd}%
\end{equation}
Here, the quantity $J(N)$ is defined as $
J(N)=\left[{1+f(N-1)\over Af} \right]^{\frac{1}{f}} $, and $k_{2}$
is a  constant given by $ k_{2}=k_{1}K^{-\frac{1-m}{4-m}}. $

From the definition of the scalar spectral index $n_{s}$, given by
$n_{s}-1=\frac{d\ln \,{\mathcal{P}_{R}}}{d\ln k}$, then
considering  Eqs. (\ref{fisol}) and (\ref{pd}), the scalar spectral
index in the weak dissipative regime results

\begin{equation}
n_{s}=1-{5-m-2f(4-m)\over Af(4-m)(B^{-1}[K\,\phi])^f} +n_2 +n_3,
\label{nss1}%
\end{equation}
where the quantities $n_2$ and $n_3$ are defined as

$$
n_{2}={1-m \over 4-m}\sqrt{{2(1-f)\over  \kappa Af}}
{(B^{-1}[K\,\phi])^{-f/2}\over \phi} \left[
1-A\left({\kappa\left(B^{-1}[K\,\phi]\right)^{2(1-f)}
\over3\,\alpha^2f^2}\right) ^{(1+\beta)} \right]^{-\beta\over
2(1+\beta)},
$$
and
$$
n_{3}=2A\beta{(3-m) \over (4-m)}{ (1-f)\left(\kappa/
3\right)^{1+\beta} \over (Af)^{3+2\beta}}
(B^{-1}[K\,\phi])^{2-3f+2\beta(1-f))} \left[
1-A\left({\kappa\left(B^{-1}[K\,\phi]\right)^{2(1-f)}
\over3\,\alpha^2f^2}\right) ^{(1+\beta)} \right]^{-1},
$$
respectively.

In fact, the scalar spectral index also can be rewritten  in
terms of the number of $e-$folds $N$. From Eqs.(\ref{N1}) and
(\ref{al}) we have that
\begin{equation}
n_{s}=1-\frac{5-m-2f(4-m)}{(4-m)[1+f(N-1)]}+n_2 +n_3, \label{nswr}%
\end{equation}
where the functions $n_2=n_2(N)$ and $n_3=n_3(N)$ now are defined
as
$$
n_{2}=K{1-m \over 4-m}\sqrt{{2(1-f)\over \kappa Af}}
{(J[N])^{-f/2}\over B[J(N)]}\left[
1-A\left({\kappa\left(J[N]\right)^{2(1-f)}
\over3\,\alpha^2f^2}\right) ^{(1+\beta)} \right]^{-\beta\over
2(1+\beta)},
$$
and
$$
n_{3}=2A\beta{(3-m) \over (4-m)}{ (1-f)\left(\kappa/
3\right)^{1+\beta} \over (Af)^{3+2\beta}}
(J[N])^{2-3f+2\beta(1-f))} \left[
1-A\left({\kappa\left(J[N]\right)^{2(1-f)}
\over3\,\alpha^2f^2}\right) ^{(1+\beta)} \right]^{-1}
$$, respectively.

On the other hand, tensor perturbations do not couple to the thermal background, so gravitational waves are only
generated by quantum fluctuations, as in standard inflation \cite{Taylor:2000ze}
\begin{equation}
{\mathcal{P}}_{g}=8\kappa\left(\frac{H}{2\pi}\right)^{2}.\label{Pg}
\end{equation}

From this spectrum, it is possible to construct a fundamental  observational
quantity, namely the tensor-to-scalar ratio
$r={\mathcal{P}}_{g}/{\mathcal{P}_{\mathcal{R}}}$.
 In this way, from Eq.(\ref{pd}) and the expression of ${\mathcal{P}}_{g}$,  the tensor-to-scalar ratio as function of the scalar field
 yields
\begin{equation}
r(\phi)={2\,\kappa\,\alpha^2 f^2 \over \pi^2\,k_{1}}
\,\phi^{\,-\frac{1-m}{4-m}} \left(B^{-1}[K\,\phi]
\right)^{-\frac{3-m}{4-m}}\left[ 1-
A\left({\kappa\left(B^{-1}[K\,\phi]\right)^{2(1-f)}
\over3\,\alpha^2f^2}\right) ^{(1+\beta)} \right]^{-
\beta(3-m)\over (1+\beta)(4-m)}
.\label{Rk}%
\end{equation}

In a similar way to the case of the scalar perturbations, the tensor-to-
scalar ratio can be expressed in terms of the number of $e-$
folds $N$, resulting
\begin{equation}
r(N) = {2\,\kappa\,\alpha^2 f^2 \over \pi^2\,k_{2}}
(B[J(N)])^{-\frac{1-m}{4-m}} \left(J[N] \right)^{-\frac{3-m}{4-m}}
\left[ 1-A\left({\kappa\left(J[N]\right)^{2(1-f)}
\over3\,\alpha^2f^2}\right) ^{(1+\beta)} \right]^{-
\beta(3-m)\over (1+\beta)(4-m)}
. \label{Rk11}%
\end{equation}
Here, we have used Eqs.(\ref{N1}) and (\ref{Rk}).

\begin{figure}[th]
{{\hspace{0cm}\includegraphics[width=3.in,angle=0,clip=true]{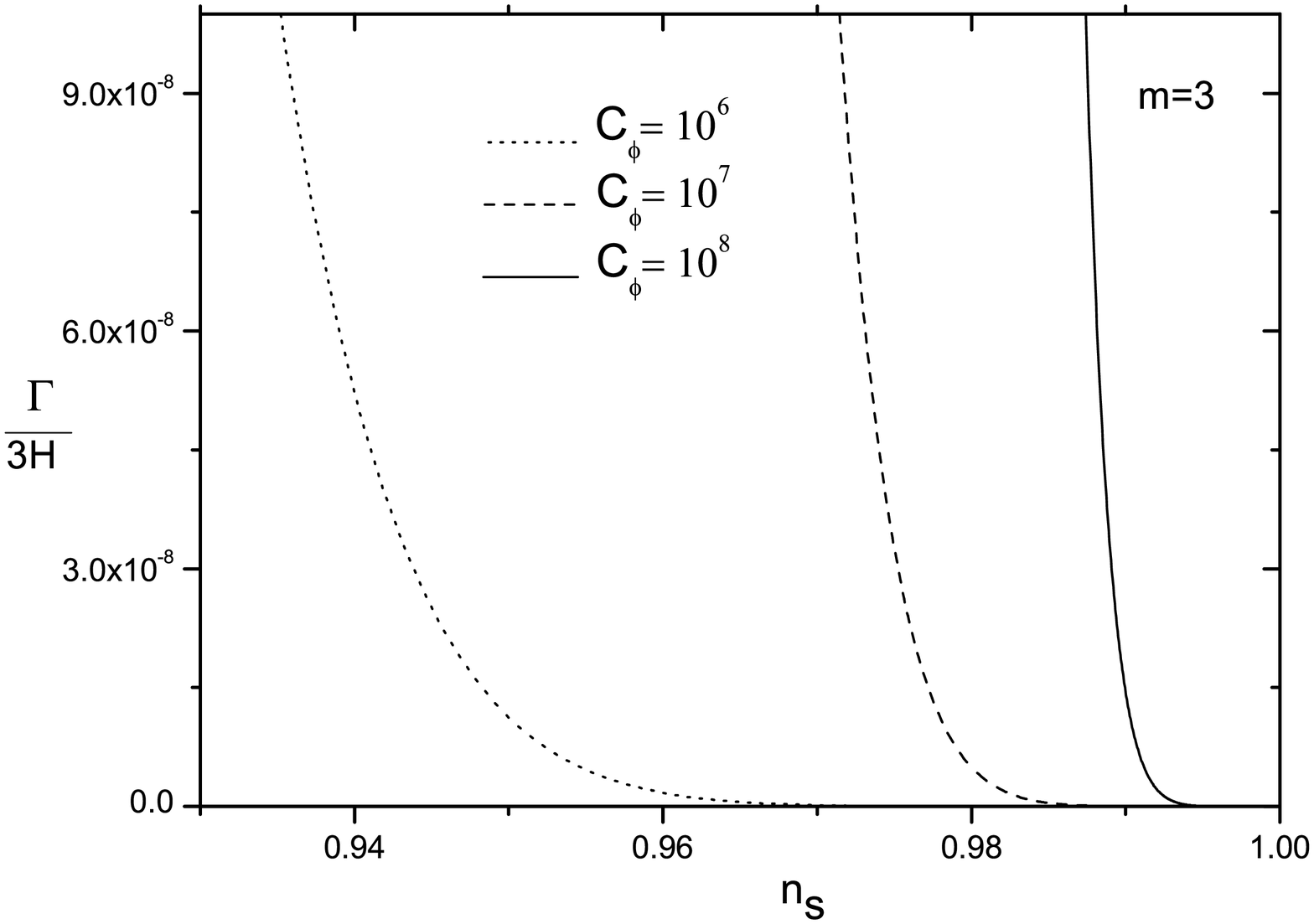}}}
{\includegraphics[width=3.in,angle=0,clip=true]{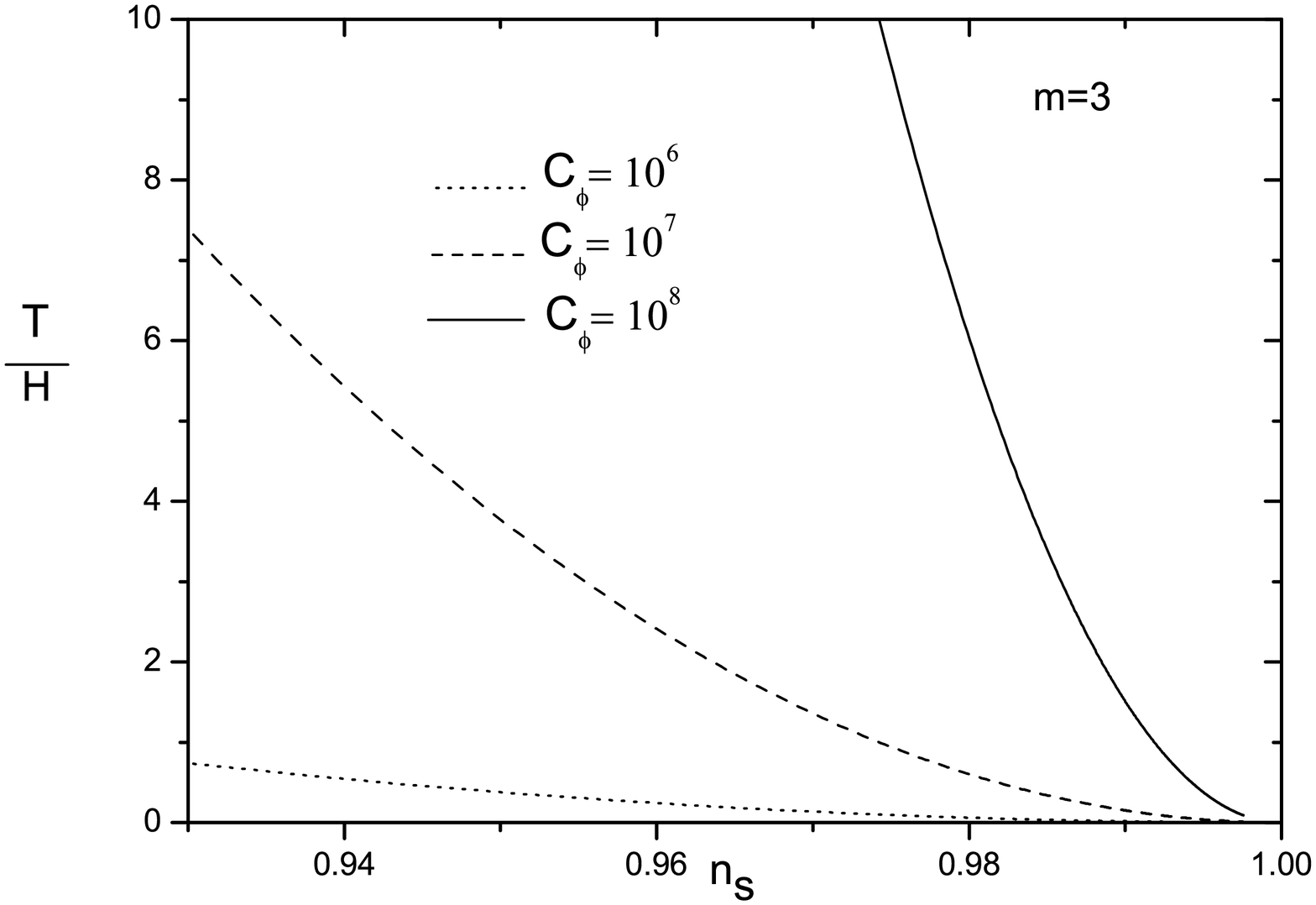}}

{\vspace{-0.5 cm}\caption{ Left panel:  ratio $\Gamma/3H$ versus
the scalar spectral index $n_s$. Right panel: ratio $T/H$ versus
the scalar spectral index $n_s$. For both panels we have
considered different values of the parameter $C_\phi$ for the
special case $m=3$ i.e., $\Gamma\propto T^3/\phi^2$, during  the
weak dissipative regime. In both panels, the dotted, dashed, and
solid lines correspond to the pairs ($\alpha=0.009$, $f=0.583$),
($\alpha=0.005$, $f=0.583$), and ($\alpha=0.002$, $f=0.582$),
respectively. In these plots we have used the values
$C_\gamma=70$, $A=0.775$, $\beta=0.00126$ and $m_p=1$ .
 \label{fig1}}}
\end{figure}

\begin{figure}[th]
{{\hspace{-1.4in}\includegraphics[width=4.2in,angle=0,clip=true]{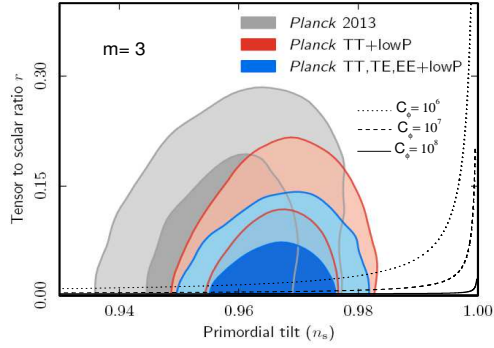}}}

{\vspace{-1.8 cm}\caption{ Evolution of the  tensor-to-scalar
ratio $r$ versus the scalar spectral index $n_s$ in the weak
dissipative regime for the special case $\Gamma\propto T^3/\phi^2$
i.e., $m=3$. Here, we have considered the two-dimensional
marginalized constraints from the new data of Planck
2015\cite{Planck2015}. In this plot we have considered   three
different values of the parameter $C_\phi$. In this panel, the
dotted, dashed,  and solid lines correspond to the pairs
($\alpha=0.009$, $f=0.583$), ($\alpha=0.005$, $f=0.583$), and
($\alpha=0.002$, $f=0.582$), respectively. As before,
 we have used the values $C_\gamma=70$,
$\rho_{Ch0}=1$, $A=0.775$, $\beta=0.00126$ and $m_p=1$.
 \label{fig2}}}
\end{figure}

\begin{figure}[th]
{{\hspace{0cm}\includegraphics[width=3.in,angle=0,clip=true]{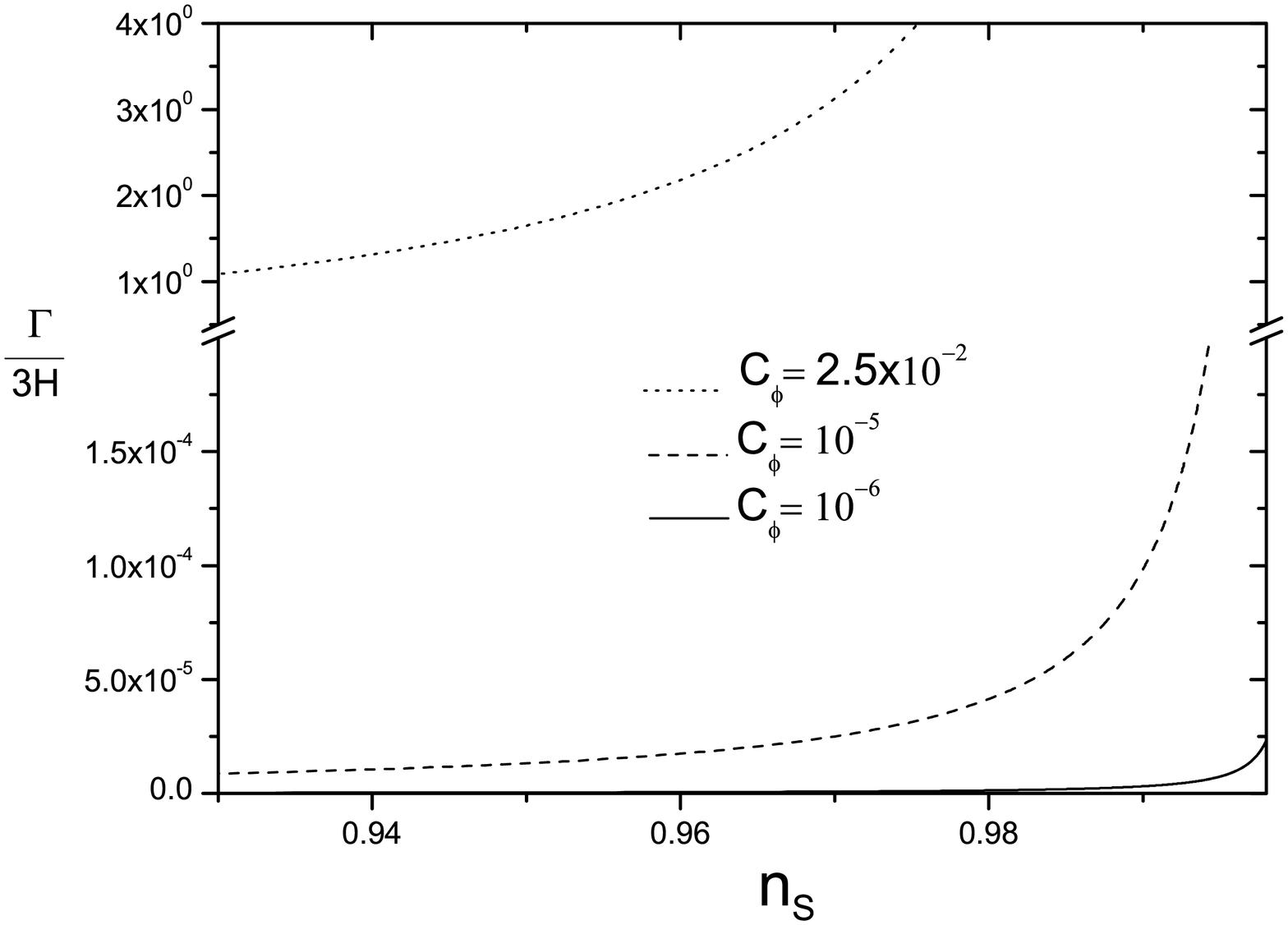}}}
{\includegraphics[width=3.in,angle=0,clip=true]{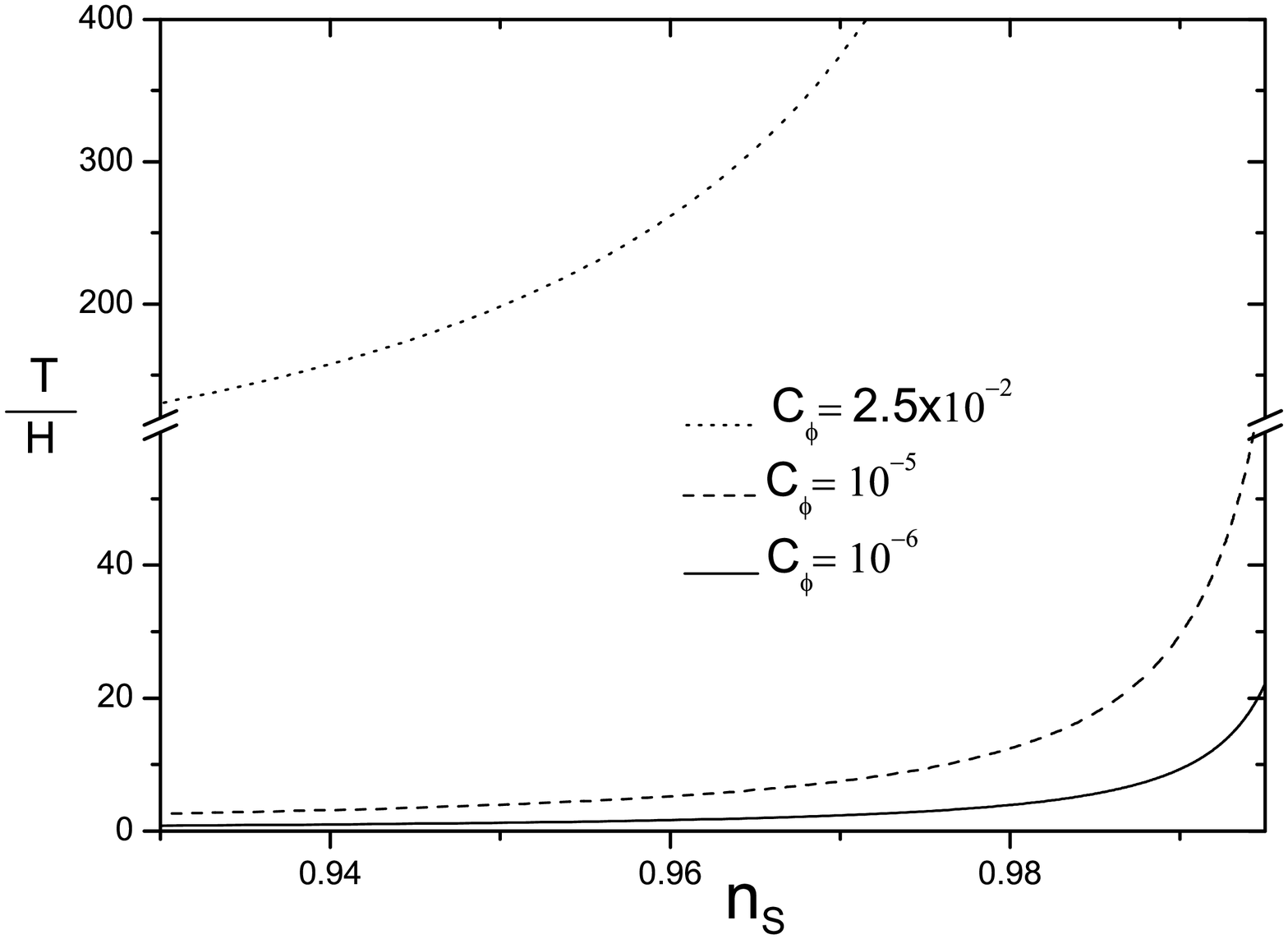}}
{\vspace{0. cm}\caption{ Evolution of the  ratio $\Gamma/3H$
versus the scalar spectral index $n_s$ (left panel) and the
evolution of the ratio $T/H$ versus  the scalar spectral index
$n_s$ (right panel) in the weak dissipative regime for the
specific value of $m=1$ ($\Gamma\propto T$), for  three different
values of the parameter $C_\phi$. In both panels, the dotted,
dashed,  and solid lines correspond to the pairs ($\alpha=0.377$,
$f=0.296$), ($\alpha=0.674$, $f=0.294$), and ($\alpha=0.798$,
$f=0.296$), respectively. Also, we have used  the values
$C_\gamma=70$, $\rho_{Ch0}=1$, $A=0.775$, $\beta=0.00126$ and
$m_p=1$ .
 \label{fig3}}}
\end{figure}

In the left and right  panels of Fig.\ref{fig1} we show the
evolution of the ratio $\Gamma/3H$
 versus the scalar spectral index  and  the evolution of the ratio $T/H$ versus the the scalar
spectral index, during the weak dissipative regime for the special
case $m=3$ i.e., $\Gamma(\phi,T)=C_\phi\,T^3/\phi^2$. In both
panels, we have considered  different values of the parameter
$C_\phi$. In fact, the left panel shows the condition $\Gamma<3H$
for the weak dissipative regime. In the right panel we show the
essential condition for warm inflation scenario to occur, given $T>H$.

In order to write down quantities that relate $\Gamma/3H$, $T/H$
and the spectral index $n_s$, we consider Eqs.(\ref{at}),
(\ref{G1}) and (\ref{fisol}), and  we obtain numerically in first
place the ratio $\Gamma/3H$ as a function of the scalar spectral
index $n_s$. Also, combining  Eqs.(\ref{at}) and (\ref{rh-1}), we
find numerically the ratio between the temperature  $T$ and the
Hubble rate $H$ as a function  of the spectral index $n_s$. In
both panels, we use the values $C_\gamma=70$, $\rho_{Ch0}=1$,
$A=0.775$, $\beta=0.00126$, see Ref.\cite{const2} and $m_p=1$.
Here we find numerically, from Eqs.(\ref{pd}) and (\ref{nswr}),
that the values $\alpha=0.009$ and $f=0.583$ correspond to the
parameter $C_\phi=10^{6}$. Here, we have used the values
${\mathcal{P}_{\mathcal{R}}}=2.43\times10^{-9}$,
 $n_s=0.97$, and the number of $e$-folds $N=60$.  In the same way, for the value of the parameter $C_\phi=10^{7}$, we find numerically
  the values  $\alpha=0.005$ and $f=0.583$. By other hand,
 for the parameter $C_\phi=10^{8}$, we find the values $\alpha=0.002$ and $f=0.582$. From the
 left panel, we obtain an upper bound for  $C_\phi<10^8$, considering  the condition for the
 weak dissipative regime $\Gamma<3H$. From the right panel
  we find  a lower bound for the parameter $C_\phi>10^6$, from the essential condition for
  warm inflation to occur, given by $T>H$.

  In Fig.\ref{fig2} we show the
  consistency relation $r=r(n_s)$ for the specific  case of
  $m=3$. Here,
  we observe  that the tensor-to-scalar ratio becomes $r\sim 0$ for the range  $10^6<C_\phi<10^8$,
 during the weak
 dissipative regime (see figure). In this form,
the range for the parameter $C_\phi$ is well corroborated from
the Planck 2015 data \cite{Planck2015}. However, we note that the
  consistency relation $r=r(n_s)$ does not impose a constraint on
  the parameter $C_\phi$ for the weak dissipative regime. In this way,
 for the specific case of $m=3$, the range of the parameter $C_\phi$ during the weak dissipative regime is given by
 $10^{6}< C_\phi < 10^{8}$.

In Fig.\ref{fig3} we show the evolution of the ratio $\Gamma/3H$
 versus the scalar spectral index (left panel) and  the evolution of the ratio $T/H$ versus the the scalar
spectral index (right panel), during the weak dissipative regime
for the special case $m=1$ i.e., $\Gamma(\phi,T)\propto \,T$. As
before, we consider Eqs.(\ref{at}), (\ref{rh-1}),
(\ref{G1})and(\ref{fisol}), and we  find numerically the ratio
$\Gamma/3H$ and the ratio between the temperature  $T$ and the
Hubble rate $H$ in terms of the scalar spectral index $n_s$, for
three different  values of the parameter $C_\phi$. Again, in both
panels, we use the values $C_\gamma=70$, $A=0.775$,
$\beta=0.00126$ and $m_p=1$. As before,  we find numerically, from
Eqs.(\ref{pd}) and (\ref{nswr}), that the values $\alpha=0.377$
and $f=0.296$ correspond to the value of the parameter
$C_\phi=0.025$. Here, again we have used the values
${\mathcal{P}_{\mathcal{R}}}=2.43\times10^{-9}$,
 $n_s=0.97$, and the number of $e$-folds $N=60$. As before, for the value $C_\phi=10^{-5}$, we
 find  numerically the values $\alpha=0.674$ and $f=0.294$, and  for $C_\phi=10^{-6}$, we obatin $\alpha=0.798$ and
 $f=0.296$. By the other hand, we study the
  consistency relation $r=r(n_s)$ for the specific  case of
  $m=1$, and we observe that the parameter $C_\phi$ is well corroborated from
  the latest data of Planck (figure not shown). Again, we note that the
ratio $\Gamma/3H<1$ gives us an upper bound for $C_\phi$, while the
condition for warm inflation $T>H$, gives us the lower bound for the
parameter $C_\phi$. In this way, for the special case in which
$m=1$, the range of the parameter $C_\phi$ during the weak
dissipative regime is given by
 $10^{-6}< C_\phi < 0.025$.

For the cases $m=0$ and $m=-1$,  and considering  the condition
for the weak dissipative regime $\Gamma<3H$, we find an upper
bound for the parameter $C_\phi$; for the case $m=0$, this bound
is found to be $C_\phi<10^{-7}$. We find numerically the values
$\alpha=0.633$ and $f=0.269$, corresponding to $C_\phi=10^{-7}$.
For the case $m=-1$, this bound is given by $C_\phi<10^{-12}$, and
for $C_\phi=10^{-12}$ we find the values $\alpha=0.817$ and
$f=0.254$. Now, from the essential condition for warm inflation to
occur $T>H$, as before, we obtain
 a lower bound for $C_\phi$; for the specific value $m=0$ i.e.,
 $\Gamma\propto\phi$ the lower bound is given by $C_\phi>10^{-12}$,
 finding numerically the values $\alpha=1.152$, $f=0.270$ for $C_\phi>10^{-12}$. Finally, for the value $m=-1$ the bound
  is given by $C_\phi>10^{-18}$. In this case, for $C_\phi=10^{-18}$, we find the values $\alpha=1.443$, $f=0.255$. Moreover, we
  observe that these values for $C_\phi$  are well corroborated from the latest
  data of Planck, considering  the  consistency relation $r=r(n_s)$ for
the cases $m=0$ and $m=-1$ (not shown). However, this consistency relation does not impose a
constraint on $C_\phi$.

It is interesting to note that the range for the parameter
$C_\phi$ for the weak dissipative regime is obtained only from the
condition for the weak dissipative regime $\Gamma<3H$, which gives us an upper bound, and the essential
condition for warm inflation to occur $T>H$, which gives us a lower bound.
We observe  that the consistency relation $r=r(n_s)$ does not
impose a constraint on $C_\phi$ for this regime.
\\
\\
\begin{table}
\centering
\begin{tabular}
[c]{|c|c|c|}\hline $\Gamma=\frac{C_{\phi}T^{m}}{\phi^{m-1}}$ &
Constraints on $\alpha$ and $f$ & Constraint on $C_{\phi}$\\\hline
$m=3$ & $%
\begin{tabular}
[c]{c}%
$0.002<\alpha<0.009$\\
$0.582<f<0.583$%
\end{tabular}
\ \ $ & $10^{6}<C_{\phi}<10^{8}$\\\hline
$m=1$ & $%
\begin{tabular}
[c]{c}%
$0.377<\alpha<0.798$\\
$0.294<f<0.296$%
\end{tabular}
\ \ $ & $10^{-6}<C_{\phi}<0.025$\\\hline
$m=0$ & $%
\begin{tabular}
[c]{c}%
$0.633<\alpha<1.152$\\
$0.269<f<0.270$%
\end{tabular}
\ \ $ & $10^{-12}<C_{\phi}<10^{-7}$\\\hline
$m=-1$ & $%
\begin{tabular}
[c]{c}%
$0.817<\alpha<1.143$\\
$0.254<f<0.255$%
\end{tabular}
\ \ $ & $10^{-18}<C_{\phi}<10^{-12}$\\\hline
\end{tabular}
\caption{Results for the constraints on the parameters $\alpha$,
$f$ and $C_\phi$ during the weak dissipative regime.} \label{T1}
\end{table}

Table \ref{T1} summarizes the constraints on the parameters
$\alpha$, $f$ and $C_\phi$, for the different values of the
parameter $m$, considering a general  form for the dissipative
coefficient $\Gamma=\Gamma(T,\phi)$, in the weak dissipative
regime. We note that these constraints on our parameters, result
as consequence of the  conditions $\Gamma<3H$ (upper bound) and
$T>H$ (lower bound). Here we have used  the values
$C_\gamma=70$, $\rho_{Ch0}=1$, $A=0.775$, $\beta=0.00126$ and
$m_p=1$.

 For the sake of numerical evaluation  let us consider
different values for the parameters $A$,  $\beta$ and $C_\phi$, but
now, the parameters $\alpha$, $f$ and $m$, are fixed. In the following,
we will find numerically the values of the parameter $A$ and
$\beta$ from Eqs.(\ref{pd}) and (\ref{nswr}), considering  the
values ${\mathcal{P}_{\mathcal{R}}}=2.43\times10^{-9}$,
 $n_s=0.97$, and the number of $e$-folds $N=60$.

\begin{figure}[th]
{{\hspace{0cm}\includegraphics[width=3.in,angle=0,clip=true]{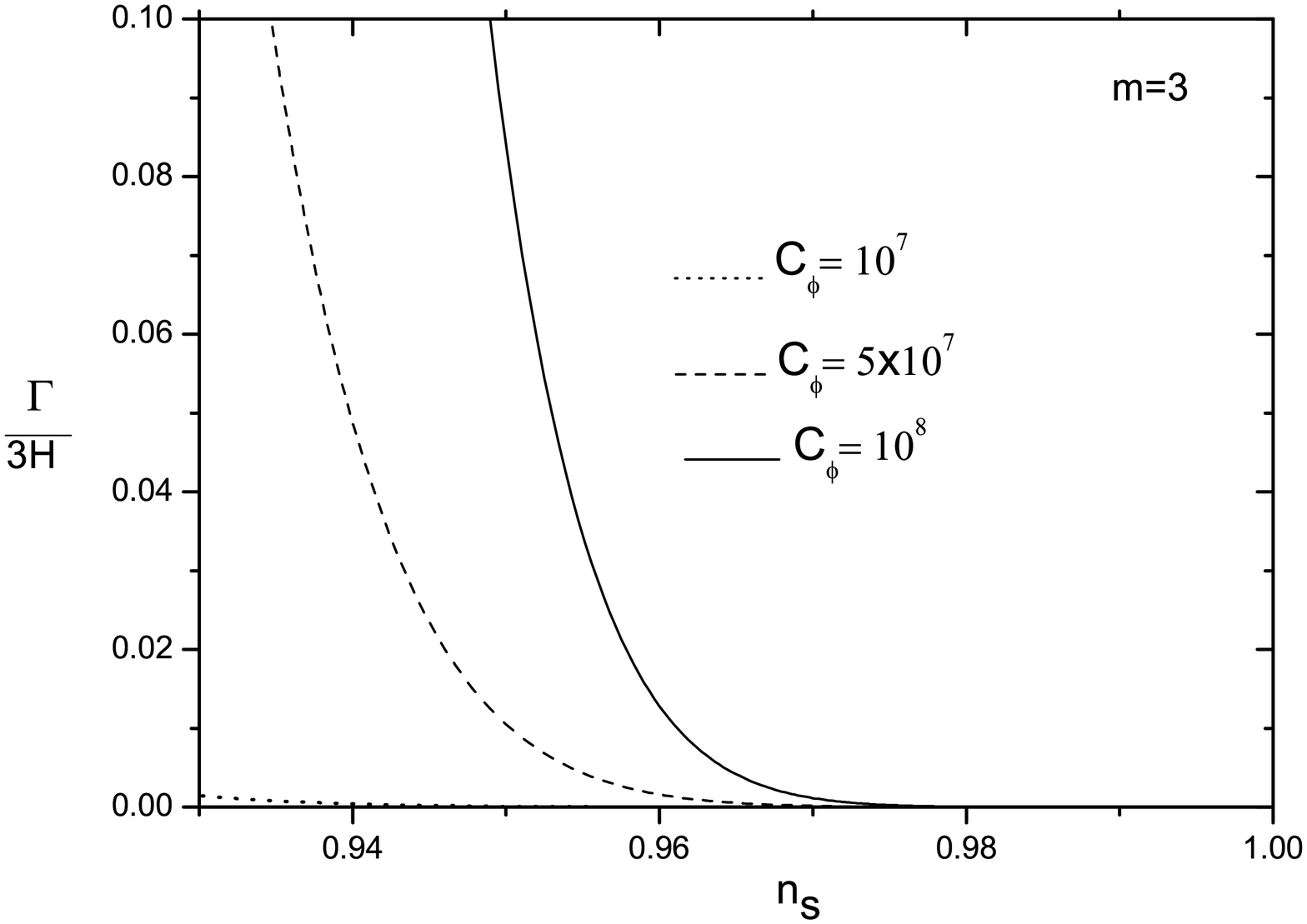}}}
{\includegraphics[width=3.in,angle=0,clip=true]{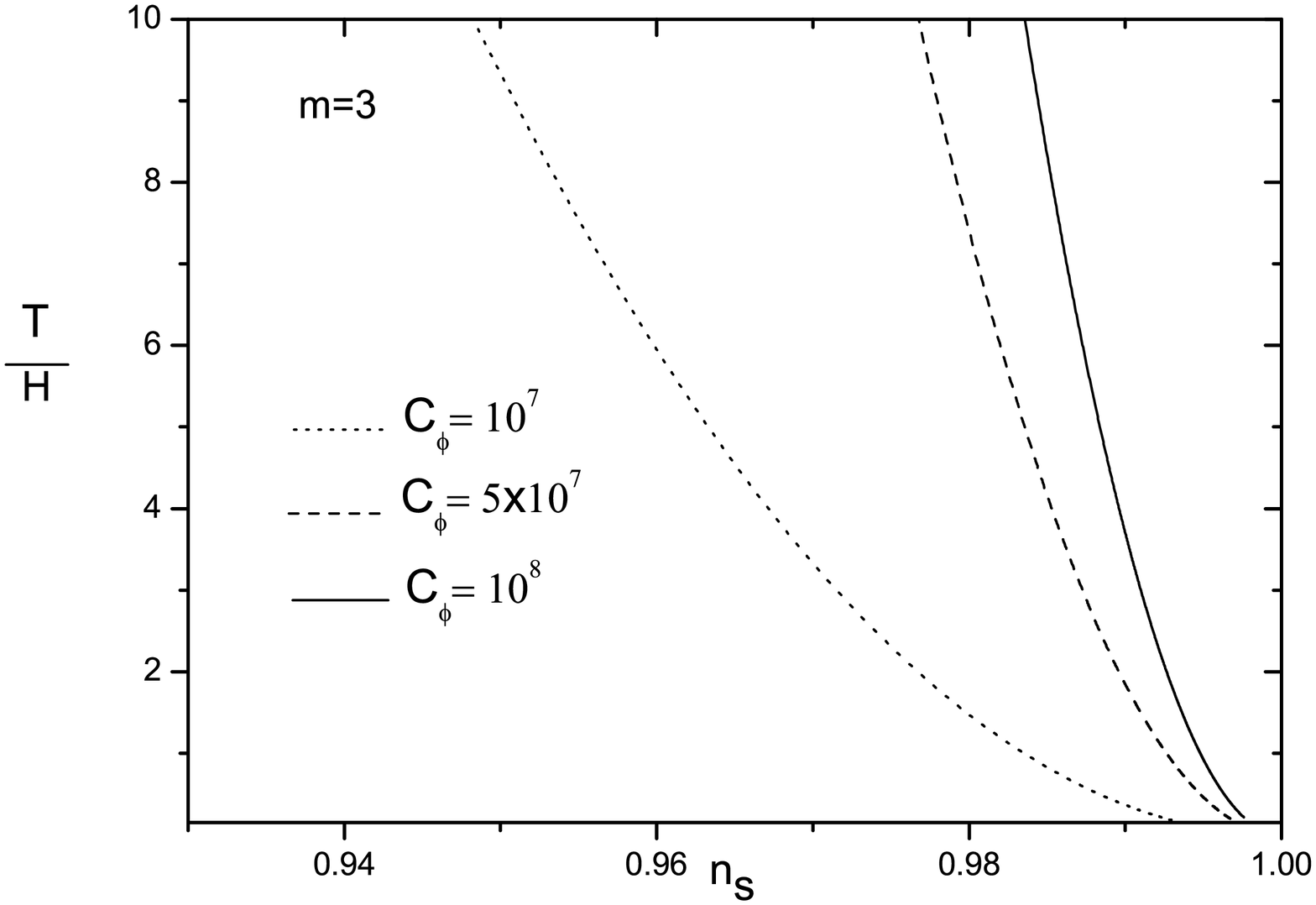}}

{\vspace{-0.5 cm}\caption{ Left panel:  ratio $\Gamma/3H$ versus
the scalar spectral index $n_s$. Right panel: ratio $T/H$ versus
the scalar spectral index $n_s$. For both panels we have
considered different values of the parameter $C_\phi$ for the
special case $m=3$ (or equivalently $\Gamma\propto T^3/\phi^2$),
during the weak dissipative regime. In both panels, the dotted,
dashed, and solid lines correspond to the pairs ($A=5.5\times
10^{-4}$, $\beta=-0.56$), ($A=2.9\times10{-3}$, $\beta=-0.61$),
and ($A=5\times10^{-3}$, $\beta=-0.62$), respectively. Now in
these plots we have used the values $C_\gamma=70$,
$\alpha=10^{-3}$, $f=0.7$ and $m_p=1$ .
 \label{fig4b}}}
\end{figure}

In the left and right  panels of Fig.\ref{fig4b} we show the
plot of  $\Gamma/3H$
 as function the scalar spectral index $n_s$ and  the plot of the ratio $T/H$ as function of the scalar
spectral index, for the weak dissipative regime, for the special
case $m=3$, considering the values $\alpha=10^{-3}$ and
$f=0.7$. As before, in both panels, we have considered different
values of the parameter $C_\phi$. Again, the left plot shows the
condition $\Gamma<3H$ for the model evolves according to
the weak dissipative regime, and in  the
right panel we show the  essential condition for warm inflation scenario to
occur, given $T>H$.

As before, for the quantities  $\Gamma/3H$, $T/H$ and the scalar spectral
index $n_s$, we take Eqs.(\ref{at}), (\ref{G1}) and (\ref{fisol}),
and we find numerically  the ratio $\Gamma/3H$ as a function of
the scalar spectral index. Also, from Eqs.(\ref{at}) and
(\ref{rh-1}), we obtain numerically the ratio $T/H$ as a function
of the spectral index $n_s$. Now we use the values $C_\gamma=70$,
$\rho_{Ch0}=1$, $\alpha=10^{-3}$, and  $f=0.7$. As before, we
obtain numerically, from Eqs.(\ref{pd}) and (\ref{nswr}), that the
values $A=5.5\times 10^{-4}$ and $\beta=-0.56$ correspond to the
parameter $C_\phi=10^{7}$. Here, we have used the observational constraints
${\mathcal{P}_{\mathcal{R}}}=2.43\times10^{-9}$,
 $n_s=0.97$, and the number of $e$-folds $N=60$.  By the other hand, for the value of the parameter $C_\phi=5\times10^{7}$, we find numerically
  the values  $A=2.9\times10^{-3}$ and $\beta=-0.61$. Likewise,
 for the parameter $C_\phi=10^{8}$, we find the values $A=5\times10^{-3}$ and $\beta=-0.62$. Here we note that from the
 left panel, we find  an upper bound for the parameter $C_\phi$, given by $C_\phi<10^8$ taking   the condition for the
 weak dissipative regime $\Gamma<3H$. Similarly, from the right panel,
  we find  a lower bound for the parameter $C_{\phi}$, given by $C_\phi>10^7$, from the essential condition for
  warm inflation to occur, given by $T>H$. Also, we note that from the
  consistency relation $r=r(n_s)$, for the specific  case of
  $m=3$, the tensor-to-scalar ratio becomes $r\sim 0$ for the range  $10^7<C_\phi<10^8$,
 during this regime (not shown). In this way,
the range for the parameter $C_\phi$ is in agreement with the
Planck 2015 results \cite{Planck2015}. As before, we observe  that
the
  consistency relation $r=r(n_s)$ does not impose any constraint on
  the parameter $C_\phi$ for the weak dissipative regime when the parameters $\alpha$ and $f$ are fixed.

  In this way,  for the special
case $m=3$, the ranges  for the parameters $C_\phi$, $A$ and
$\beta$ during the weak dissipative regime are  given by
$10^7<C_\phi<10^8$, $5.5\times 10^{-4}<A<5\times 10^{-3}$ and
$-0.62<\beta<-0.55$, respectively. We note that in the
representation  of the GCG as a variant of gravity,  our analysis
favors negative values for the parameter $\beta$.

For the case  $m=1$,  and considering that our model evolves according to the weak
dissipative regime, i.e., $\Gamma<3H$, we find an upper bound for the
parameter $C_\phi$. Analogously as before, and when the parameters $\alpha$ and $f$ are fixed to be
$\alpha=10^{-3}$ and $f=0.7$, respectively, we numerically find that the values
$A=154.7$ and $\beta=-1.3$ correspond to
$C_\phi=1.1\times10^{-2}$.  Now, from the essential condition for
warm inflation to occur $T>H$,  we obtain
 a lower bound for $C_\phi$, given by $C_\phi>3.6\times10^{-5}$, and
 numerically find that the values $A=2995$, and $\beta=-1.4$, correspond to $C_\phi=3.6\times10^{-5}$. We
  observe that these values for $C_\phi$  are well corroborated from the
   Planck 2015 results. Moreover, the tensor-to-scalar ratio becomes $r\sim 0$ (not shown), and as before the $r-n_s$
plane does not impose any constraint on $C_\phi$. In this way,
for the special case $m=1$, the allowed ranges for the parameters
$C_\phi$, $A$, and $\beta$ for the weak dissipative regime are
given by $3.6\times10^{-5}<C_\phi<1.1\times10^{-2}$,
$154.7<A<2995$ and $-1.4<\beta<-1.3$, respectively. We note
that our analysis favors negative values for $\beta$

For the cases in which $\Gamma\propto \phi$ (or equivalently
$m=0$) and $\Gamma\propto \phi^2/T$ (or equivalently $m=-1$), we
find  that these models do not work in the weak dissipative
regime, since the scalar spectral index $n_s>1$, and then these
cases are disproved from the observational data.

Table \ref{T1b} summarizes the constraints on the parameters $A$,
$\beta$ and $C_\phi$, for the different values of the parameter
$m$, considering a general  form for the parameter
$\Gamma=\Gamma(T,\phi)$, in the weak dissipative regime. As
before, these constraints result as a consequence of
the  conditions $\Gamma<3H$ (upper bound) and $T>H$ (lower bound).
 Here, we have fixed the values $C_\gamma=70$, $\rho_{Ch0}=1$,
$\alpha=10^{-3}$, $f=0.7$ and $m_p=1$.
\\
\\
\begin{table}
\centering
\begin{tabular}
[c]{|c|c|c|}\hline $\Gamma=\frac{C_{\phi}T^{m}}{\phi^{m-1}}$ &
Constraints on $A$ and $\beta$ & Constraint on $C_{\phi}$\\\hline
$m=3$ & $%
\begin{tabular}
[c]{c}%
$5.5\times10^{-4}<A<5\times10^{-3}$\\
$-0.62<\beta<-0.55$%
\end{tabular}
\ \ $ & $10^{7}<C_{\phi}<10^{8}$\\\hline $m=1$ &
\begin{tabular}
[c]{c}%
$155<A<2995$\\
$-1.4<\beta<-1.3$%
\end{tabular}
& $3.6\times10^{-5}<C_{\phi}<1.1\times 10^{-2}$\\\hline $m=0$ &
The model does not work $(n_{s}>1)$ & --\\\hline $m=-1$ & The
model does not work $(n_{s}>1)$ & --\\\hline
\end{tabular}
\caption{Results for the constraints on the parameters $A$,
$\beta$ and $C_\phi$ during the weak dissipative regime.}
\label{T1b}
\end{table}

\section{ The strong dissipative regime.\label{subsection2}}

In this section  we analyze  the inflationary dynamics of our
Chaplygin warm model in the strong dissipative regime
 $\Gamma>3H$.
 By using Eqs. (\ref{inf3}) and
(\ref{G1}),  we obtain  the solution  of the scalar field
$\phi(t)$, in terms of the cosmic time. Here, we study  the
solution for the scalar field for two different  values of the
parameter $m$,
 namely the cases $m=3$
and $m\neq3$. For the specific  case $m=3$, the solution $\phi(t)$
is found to be
\begin{equation}
\phi(t)-\phi_0=\exp\left[\frac{\tilde{B}[t]}{\tilde{K}}\right], \label{phisf}%
\end{equation}
where $\phi(t=0)=\phi_{0}$ is an integration constant   and the
quantity  $\tilde{K}$ is a the constant defined as
$$
\tilde{K}\equiv 2^{7\over
8}\,{(1+\beta)C_{\phi}^{1/2}\over(4C_{\gamma})^{3/8}} \,
\,{\left(\kappa/3\right)^{{1\over 8}+\frac{2+5f}{16(1-f)}}\over
(\alpha\,f)^{{5\over 8}+\frac{2+5f}{16(1-f)}}} (1-f)^{7\over
8}\,A^{\frac{2+5f}{16(1+\beta)(1-f)}}, \label{Ktilde}
$$
and the function $\tilde{B}[t]$ is  given by
\begin{equation}
\tilde{B}[t]\equiv B\left[A\,\left({\kappa\over
3\alpha^2f^2}\right)^{1+\beta}\,t^{2(1+\beta)(1-f)};
\,\frac{2+5f}{16(1+\beta)(1-f)},\,\frac{8+7\beta}{8(1+\beta)}\right],\label{Betatilde}
\end{equation}
and this function corresponds to the incomplete beta function, see
Ref.\cite{Libro}.

For the specific case in which  $m\neq3$, the
solution for the scalar field is found to be
\begin{equation}
\varphi(t)-\varphi_0=\frac{\tilde{B}_{m}[t]}{\tilde{K}_{m}}, \label{phisfm}%
\end{equation}
where the new scalar field $\varphi$ is defined as
$\varphi(t)=\frac{2}{3-m}\phi(t)^{\frac{2}{3-m}}$. Again,
$\varphi_0$ corresponds to an  integration constant, that can be
assumed $\varphi_0=0$. Also,  $\tilde{K}_m$ is a  constant defined
as $\tilde{K}_m\equiv 2^{4+m\over
8}\,{(1+\beta)C_{\phi}^{1/2}\over(4C_{\gamma})^{m/8}} \,
\,{\left(\kappa/3\right)^{{4-m\over
8}+\frac{m(2-f)-4(1-2f)}{16(1-f)}}\over(\alpha\,f)^{{8-m\over
8}+\frac{m(2-f)-4(1-2f)}{16(1-f)}}} (1-f)^{4+m\over
8}\,A^{\frac{m(2-f)-4(1-2f)}{16(1+\beta)(1-f)}} $. The function
$\tilde{B}_m[t]$ in Eq.(\ref{phisfm}), for the specific case
$m\neq3$, also corresponds to the incomplete beta function, given
by
\begin{equation}
\tilde{B}_m[t]\equiv B\left[A\,\left({\kappa\over
3\alpha^2f^2}\right)^{1+\beta}\,t^{2(1+\beta)(1-f)};
\,\frac{m(2-f)-4(1-2f)}{16(1+\beta)(1-f)},\,\frac{8+\beta(4+m)}{8(1+\beta)}\right]
.\label{Betatildem}
\end{equation}

Considering  Eqs.(\ref{at}), (\ref{phisf}) and (\ref{phisfm}), the
Hubble rate as function of the scalar field may be be written as
\begin{equation}
H(\phi)=\frac{Af}{(\tilde{B}^{-1}[\tilde{K}\ln
\phi])^{1-f}},\,\,\;\;\;\text{for}\,\;\;\;m=3,\label{Hf3}
\end{equation}
and
\begin{equation}
H(\varphi)=\frac{Af}{(\tilde{B}_m^{-1}[\tilde{K}_m\varphi])^{1-f}},\;\;\;
\,\,\;\text{for}\,\;\;m\neq 3.
\end{equation}

From Eq.(\ref{pot}), we find that the effective scalar potential
$V(\phi)$ (or equivalently $V(\varphi)$), under the
slow-roll approximation, obtaining
\begin{equation}
V(\phi)\simeq \left[\left({3\,\alpha^2f^2\over
\kappa\left(\tilde{B}^{-1}[\tilde{K}\ln
\phi]\right)^{2(1-f)}}\right)^{1+\beta} -A\right]^{1\over
1+\beta},
\label{potm3}%
\end{equation}
for the specific case  $m=3$, and we obtain
\begin{equation}
V(\varphi)\simeq \left[\left({3\,\alpha^2f^2\over
\kappa\left(\tilde{B}_m^{-1}[\tilde{K}_m\varphi]\right)^{2(1-f)}}\right)^{1+\beta}
-A\right]^{1\over 1+\beta},
\label{potmd3}%
\end{equation}
for the case $m\neq3$.

Now combining Eqs. (\ref{G1}), (\ref{phisf}), and (\ref{phisfm}),
the dissipative coefficient $\Gamma$ as function of the scalar
field results
  \begin{equation}
\Gamma(\phi)=\delta
\phi^{-2}(\tilde{B}^{-1}[\tilde{K}\ln\phi])^{-\frac{3(2-f)}{4}}\left[1-A
\left({\kappa\left(\tilde{B}^{-1}[\tilde{K}\ln\phi]\right)^{2(1-f)}
\over3\,\alpha^2f^2}\right) ^{(1+\beta)}\right]
^{-\frac{3\beta}{4(1+\beta)}}, \label{gamma3}%
\end{equation}
for the case $m=3$. Here $\delta$ is a  constant and   is given by
 $\delta=C_{\phi}\left[\frac{Af(1-f)}{2\kappa C_{\gamma}}\right]^{3/4}$.  For the special case in which  $m\neq3$
 we find that the dissipative coefficient becomes
\begin{equation}
\Gamma(\phi)=\delta_m
\phi^{1-m}(\tilde{B}_m^{-1}[\tilde{K}_m\varphi])^{-\frac{m(2-f)}
{4}}\left[1-A\left({\kappa\left(\tilde{B}_m^{-1}[\tilde{K}_m\varphi]\right)^{2(1-f)}
\over3\,\alpha^2f^2}\right) ^{(1+\beta)}\right]
^{-\frac{\beta m}{4(1+\beta)}}, \label{gammasm}%
\end{equation}
where  $\delta_m=C_{\phi}\left[\frac{Af(1-f)}{2\kappa
C_{\gamma}}\right]^{m/4}$ is a constant.

During  the strong dissipative regime,
 the dimensionless slow-roll parameter $\varepsilon$ is defined as  $
\varepsilon=-\frac{\dot{H}}{H^{2}}=\frac{1-f}{Af(\tilde{B}^{-1}[\tilde{K}\ln\phi])
^f}$, for the specific case of $m=3$ and  for the case $m\neq3$,
this parameter becomes
$\varepsilon=\frac{1-f}{Af(\tilde{B}_m^{-1}[\tilde{K}_
m\varphi])^f}$. Analogous to the case of the weak dissipative
regime,  if $\ddot{a}>0$, then the scalar field
$\phi>\exp[\frac{1}{\tilde{K}}\tilde{B}[(\frac{1-f}{Af})^{1/f}]]$
 for  $m=3$, and for the case $m\neq3$  results
$\varphi>\frac{1}{\tilde{K}_m}\tilde{B}_m[(\frac{1-f}{Af})^{1/f}]$
. As before, the value of the scalar field at the beginning of inflation
 is
$\phi_1=\exp[\frac{1}{\tilde{K}}\tilde{B}[(\frac{1-f}{Af})^{1/f}]]$,
for the specific value of $m=3$, and for the special case
 $m\neq3$ we get
$\varphi_1=\frac{1}{\tilde{K}_m}\tilde{B}_m[(\frac{1-f}{Af})^{1/f}]$.

 In relation to the number of $e$-folds $N$ in the strong regime,
 we find that combining
   Eqs.(\ref{at}), (\ref{phisf}), and (\ref{phisfm}) yields
\begin{equation}
N=\int_{t_{1}}^{t_{2}}\,H\,dt=\alpha[(\tilde{B}^{-1}[\tilde{K}\ln\phi_2])^f-(\tilde{B}^{-1}[\tilde{K}\ln\phi_1])^f],\,\,\;\;\;\mbox{for}\,\;\;\,m=3,\label{Nsr3}%
\end{equation}
and
\begin{equation}
N=\alpha[(\tilde{B}_m^{-1}[\tilde{K}_m\varphi_2])^f-(\tilde{B}_
m^{-1}[\tilde{K}_m\varphi_1])^f],\;\;\; \,\,\mbox{for}\,\,\;\;\;m\neq3.\label{Nsrm}%
\end{equation}

Now we will study
the cosmological perturbations in the strong regime
$R=\Gamma/3H>1$. Following Ref.\cite{warm}, the fluctuation
$\delta\phi^2$ in the strong dissipative regime is found to be
$\delta\phi ^{2}\simeq\frac{k_{F}T}{2\pi^{2}}$, where the
function $k_{F}$ corresponds to the freeze-out wave-number,
defined as $k_{F}=\sqrt{\Gamma H}=H\sqrt{3R}>H$. In this form,
the power spectrum of the
scalar perturbation $P_{\mathcal{R}}$, considering
Eqs.(\ref{at}), (\ref{rh-1}) and (\ref{G1}) results
\begin{equation}
P_{\mathcal{R}}\simeq\frac{H^{\frac{5}{2}}\Gamma^{\frac{1}{2}}T}{2\pi^{2}%
\dot{\phi}^{2}}=\frac{\kappa}{12\pi^2}C_{\phi}^{3/2}\left(\frac{3}{2\kappa C_{\gamma}}
\right)^{\frac{3m+2}{8}}\phi^{\frac{3(1-m)}{2}}H^{3/2}(-\dot{H})^{\frac{3m-6}{8}}\left[1-
A\left({3\,H^2 \over \kappa}\right)^{-(1+\beta)}\right]^{-\frac{\beta (3m-6)}{8(1+\beta)}}.\label{Prsr}%
\end{equation}

Also, the power spectrum $\mathcal{P}_{\mathcal{R}}$ may be
express in terms of the scalar field $\phi$. From Eqs. (\ref{at}), (\ref{phisf}),
(\ref{phisfm}) and  (\ref{Prsr}), we obtain that the scalar power
spectrum becomes
\begin{equation}
\mathcal{P}_{\mathcal{R}}=k(\tilde{B}^{-1}[\tilde{K}\ln\phi])^
{\frac{3(5f-6)}{8}}\phi^{-3}\left[1-A\left({\kappa\left(\tilde{B}^{-1}[\tilde{K}\ln\phi]\right)^{2(1-f)}
\over3\,\alpha^2f^2}\right) ^{(1+\beta)}\right]
^{-\frac{3\beta}{8(1+\beta)}},\label{Prf3}
\end{equation}
for the special case of $m=3$. Here
 $k$ is a constant and is defined as $k=\frac{\kappa}{12\pi^2}C_{\phi}^{3/2}
 \left(\frac{3}{2\kappa C_{\gamma}}\right)^{11/8}(Af)^{15/8}(1-f)^{3/8}$.
For the  case of $m\neq3$, we find that the power spectrum becomes
written as
 \begin{equation}
\mathcal{P}_{\mathcal{R}}=k_m(\tilde{B}_m^{-1}[\tilde{K}_m\varphi])^
{\frac{3[f(m+2)-2m]}{8}}\phi^{\frac{3}{2}(1-m)}\left[1-A\left({\kappa\left(\tilde{B}_m^{-1}[\tilde{K}_m\varphi]\right)^{2(1-f)}
\over3\,\alpha^2f^2}\right) ^{(1+\beta)}\right] ^{-\frac{\beta
(3m-6)}{8(1+\beta)}},\label{Prfm}
\end{equation}
where the constant $k_m$,  is given by
$k_m=\frac{\kappa}{12\pi^2}C_{\phi}^{3/2}\left(\frac{3}{2\kappa
C_{\gamma}}\right)^{\frac{3m+2}{8}}(Af)^{\frac{3m+6}{8}}(1-f)^{\frac{3m-6}{8}}$.

In similar way, the scalar power spectrum  can be
expressed in terms of the number of $e$-folds $N$. Combining
 Eqs.(\ref{Nsr3}) and (\ref{Nsrm}) in (\ref{Prf3}) and (\ref{Prfm})
 we have

\begin{equation}
\mathcal{P}_{\mathcal{R}}=k(J[N])^{\frac{3(5f-6)}{8}}\exp\left
(-\frac{3}{\tilde{K}}\tilde{B}[J[N]]\right)\left[
1-A\left({\kappa\left(J[N]\right)^{2(1-f)}
\over3\,\alpha^2f^2}\right) ^{(1+\beta)}
\right]^{-\frac{3\beta}{8(1+\beta)}},\label{Prf3N}
\end{equation}
for the particular case of $m=3$. For the specific case $m\neq3$
we obtain
\begin{equation}
\mathcal{P}_{\mathcal{R}}=\tilde{\gamma}_m(J[N])^{\frac{3[f(2+m)-2m]}{8}}
(\tilde{B}_m[J[N]])^{\frac{3(1-m)}{3-m}}\left[
1-A\left({\kappa\left(J[N]\right)^{2(1-f)}
\over3\,\alpha^2f^2}\right) ^{(1+\beta)} \right]^{-\frac{\beta
(3m-6)}{8(1+\beta)}},\label{PrfmN}
\end{equation}
where the constant $\tilde{\gamma}_m$  is defined as
$\tilde{\gamma}_m=k_m\left(\frac{2\tilde{K}_m}{3-m}\right)^{-\frac{3(1-m)}{3-m}}$.

Now combining  Eqs. (\ref{Prf3}) and (\ref{Prfm}), we find that
that the scalar spectral index $n_s$ is found to be
\begin{equation}
n_s=1+\frac{3(5f-6)}{8Af}(\tilde{B}^{-1}[\tilde{K}\ln\phi])^{-f}+
n_1+n_2,\label{nsf3}
\end{equation}
for the value $m=3$. Here the quantities $n_1$ and $n_2$ are
defined as
$$
n_1=-3\left(\frac{6}{\kappa}\right)^{1/2}\frac{1}{C_{\phi}^{1/2}}\left(\frac{3}{2\kappa
C_{\gamma}}\right)^{-3/8}(Af)^{-3/8}\times
(1-f)^{1/8}(\tilde{B}^{-1}[\tilde{K}\ln\phi])^{\frac{1}{8}(2-3f)}\,\;\;\times
$$
$$
\left[1-A\left({\kappa\left(\tilde{B}^{-1}[\tilde{K}\ln\phi]\right)^{2(1-f)}
\over3\,\alpha^2f^2}\right) ^{(1+\beta)}\right]
^{-\frac{3\beta}{8(1+\beta)}},
$$ and
$$n_2=\frac{3^{-\beta}}{4}\kappa^{1+\beta}A \beta (1-f)(Af)^{-(3+2\beta)}(\tilde{B}^{-1}[\tilde{K}\ln\phi])
^{-f(3+2\beta)+2(1+\beta)}\;\;\;\;\times
$$
$$\left[1-A\left({\kappa\left(\tilde{B}^{-1}[\tilde{K}\ln\phi]\right)^{2(1-f)}
\over3\,\alpha^2f^2}\right) ^{(1+\beta)}\right] ^{-1},$$
 respectively. For the case
 $m\neq3$, we find that the scalar spectral index  results

\begin{equation}
n_s=1+\frac{3[f(m+2)-2m]}{8Af}(\tilde{B}_m^{-1}[\tilde{K}_m\varphi])^{-f}+n_{1_{m}}+n_{2_{m}},\label{nsfm}
\end{equation}
where   the  functions $n_{1_{m}}$ and $n_{2_{m}}$ are given by
$n_{1_{m}}=\frac{3(1-m)}{2}\left(\frac{6}{\kappa}\right)^{1/2}\left(\frac{3}{2\kappa
C_{\gamma}}\right)^{-m/8}\\\times\frac{1}{C_{\phi}^{1/2}}(Af)^{-m/8}(1-f)^{\frac{4-m}{8}}(\tilde{B}_m^{-1}[\tilde{K}_m\varphi])^{-\frac{[4+m(f-2)]}{8}}\phi^{\frac{m-3}{2}}
\left[1-A\left({\kappa\left(\tilde{B}_m^{-1}[\tilde{K}_m\varphi]\right)^{2(1-f)}
\over3\,\alpha^2f^2}\right) ^{(1+\beta)}\right] ^{\frac{\beta
(m-4)}{8(1+\beta)}},$ and
$n_{2_{m}}=\frac{(3m-6)}{4}\left(\frac{\kappa}{3}\right)^{1+\beta}A
\beta
(Af)^{-(3+2\beta)}(\tilde{B}_m^{-1}[\tilde{K}_m\varphi])^{-f(3+2\beta)+2(1+\beta)}\left[1-A\left({\kappa\left(\tilde{B}_m^{-1}[\tilde{K}_m\varphi]\right)^{2(1-f)}
\over3\,\alpha^2f^2}\right) ^{(1+\beta)}\right] ^{-1}$.

 Analogously as before, we may express the scalar spectral index $n_s$  in terms of the
number of $e$-folds $N$. Considering
Eqs.(\ref{Nsr3}), (\ref{Nsrm}), (\ref{nsf3}) and (\ref{nsfm}) we
obtain

\begin{equation}
n_s=1+\frac{3(5f-6)}{8Af}(J[N])^{-f}+n_1+n_2,\label{nsf3N}
\end{equation}
for the  case of $m=3$. Here the functions   $n_1$ and $n_2$
 are given by
$$
  n_1(J[N])=-3\left(\frac{6}{\kappa}\right)^{1/2}\frac{1}{C_{\phi}^{1/2}}\left(\frac{3}{2\kappa C_{\gamma}}\right)^{-3/8}(Af)^{-3/8}
  (1-f)^{1/8}(J[N])^{\frac{1}{8}(2-3f)}\\\times
$$
$$  \left[ 1-A\left({\kappa\left(J[N]\right)^{2(1-f)}
\over3\,\alpha^2f^2}\right) ^{(1+\beta)}
\right]^{-\frac{3\beta}{8(1+\beta)}},
$$
 and
$$
n_2=\frac{3^{-\beta}}{4}\kappa^{1+\beta}A \beta
(1-f)(Af)^{-(3+2\beta)}\\\times(J[N])
^{-f(3+2\beta)+2(1+\beta)}\left[1-A\left({\kappa\left(J[N]\right)^{2(1-f)}
\over3\,\alpha^2f^2}\right) ^{(1+\beta)}\right] ^{-1},
$$ respectively. For the case  $m\neq3$,
the scalar spectral index in terms of $N$ becomes
\begin{equation}
n_s=1+\frac{3[f(m+2)-2m]}{8Af}(J[N])^{-f}+n_{1_{m}}+n_{2_{m}},\label{nsfmN}
\end{equation}
where the quantities $n_{1_{m}}$ and $n_{2_{m}}$ are defined as

$$
n_{1_{m}}(J[N])=\frac{6(1-m)}{3-m}(1-f)(1+\beta)\left(\frac{\kappa
A^{\frac{1}{1+\beta}}}{3}\right)^{\frac{m(2-f)-4(1-2f)}{16(1-f)}}(Af)^{-\frac{1}{8}[4+m(2-f)]}\\\times
$$
$$
(J[N])^{-\frac{1}{8}[4+m(2-f)]}(\tilde{B}_m[J[N]])^{-1}\left[1-A\left({\kappa\left(J[N]\right)^{2(1-f)}
\over3\,\alpha^2f^2}\right) ^{(1+\beta)}\right]
^{\frac{\beta(m-4)}{8(1+\beta)}},$$
 and
$$n_{2_{m}}=\frac{(3m-6)}{4}\left(\frac{\kappa}{3}\right)^{1+\beta}A
\beta
(Af)^{-(3+2\beta)}(J[N])^{-f(3+2\beta)+2(1+\beta)}\left[1-A\left({\kappa\left(J[N]\right)^{2(1-f)}
\over3\,\alpha^2f^2}\right) ^{(1+\beta)}\right] ^{-1}.$$

Also, we find that the tensor-to-scalar-ratio $r$  in terms of the
scalar field may be written as
\begin{equation}
r=\frac{2\kappa}{\pi^2
k}(Af)^2(\tilde{B}^{-1}[\tilde{K}\ln\phi])^{\frac{(f+2)}{8}}
\phi^3\left[1-A\left({\kappa\left(\tilde{B}^{-1}[\tilde{K}\ln\phi]\right)^{2(1-f)}
\over3\,\alpha^2f^2}\right) ^{(1+\beta)}\right]
^{\frac{3\beta}{8(1+\beta)}},\label{rs3f}
\end{equation}
for the specific  case $m=3$ and
$$
r=\frac{2\kappa}{\pi^2 k_m}(Af)^2(\tilde{B}_m^{-1}[\tilde{K}_m\varphi])^{
\frac{1}{8}[6m+f(10-3m)-16]}\phi^{\frac{3}{2}(m-1)}\,\,\;\times
$$
\begin{equation}
\left[1-A\left({\kappa\left(\tilde{B}_m^{-1}[\tilde{K}_m\varphi]\right)^{2(1-f)}
\over3\,\alpha^2f^2}\right) ^{(1+\beta)}\right]
^{\frac{3\beta(m-2)}{8(1+\beta)}},\label{rsmf}
\end{equation}
for the case of $m\neq 3$.

Finally, the tensor-to-scalar ratio $r$ in terms of the
 number of $e$-folds $N$, from  Eqs.(\ref{Nsr3}) and
  (\ref{rs3f}), becomes
\begin{equation}
r=\frac{2\kappa}{\pi^2
k}(Af)^2(J[N])^{\frac{(f+2)}{8}}\exp\left[3\frac{\tilde{B}[J[N]]}
{\tilde{K}}\right]\left[1-A\left({\kappa\left(J[N]\right)^{2(1-f)}
\over3\,\alpha^2f^2}\right) ^{(1+\beta)}\right]
^{\frac{3\beta}{8(1+\beta)}},\label{rs3N}
\end{equation}
for the special case $m=3$, and  from Eqs.(\ref{Nsrm}) and
(\ref{rsmf}), the tensor to scalar ratio $r=r(N)$ becomes
$$
r=\frac{2\kappa}{\pi^2 k_m}(Af)^2(J[N])^{\frac{1}{8}[6m+f(10-3m)-16]}
\left(\frac{3-m}{2}\frac{\tilde{B}_m[J[N]]}{\tilde{K}_m}\right)^{\frac{3(m-1)}
{3-m}}\,\,\times
$$
\begin{equation}
\left[1-A\left({\kappa\left(J[N]\right)^{2(1-f)}
\over3\,\alpha^2f^2}\right) ^{(1+\beta)}\right]
^{\frac{3\beta(m-2)}{8(1+\beta)}},\label{rsmN}
\end{equation}
for the case of $m\neq3$.

\begin{figure}[th]
{{\hspace{0cm}\includegraphics[width=3.in,angle=0,clip=true]{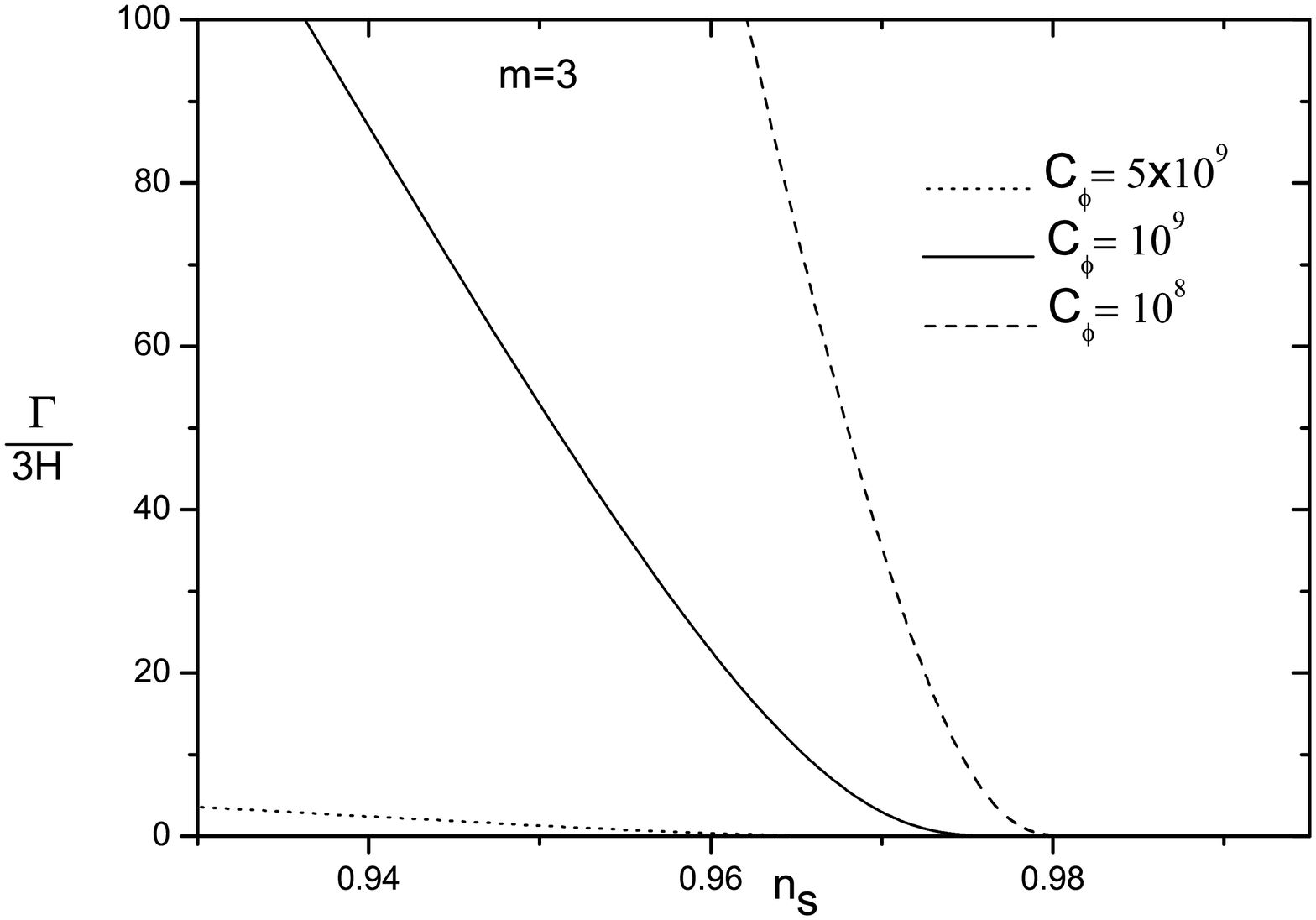}}}
{\includegraphics[width=3.in,angle=0,clip=true]{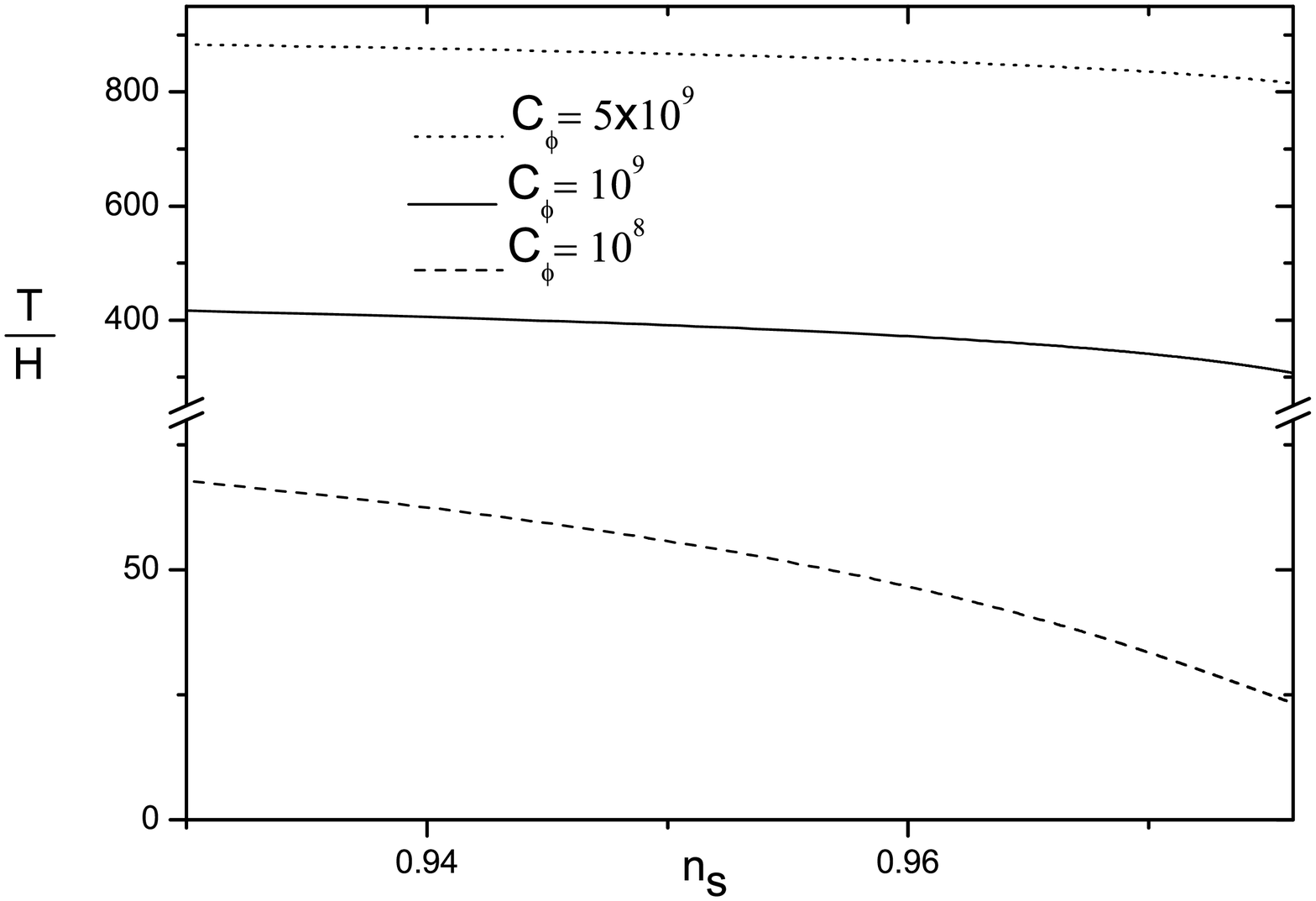}}

{\vspace{-0.5 cm}\caption{ Left panel:  ratio $\Gamma/3H$ versus
the scalar spectral index $n_s$. Right panel: ratio $T/H$ versus
the scalar spectral index $n_s$. For both panels we have used
different values of the parameter $C_\phi$, for the special case
$m=3$ i.e., $\Gamma\propto T^3/\phi^2$ during  the strong
dissipative regime. Also, in both panels, the dotted, solid, and
dashed lines correspond to the pairs ($\alpha=1.46\times10^{-5}$,
$f=0.703$), ($\alpha=6.91\times10^{-6}$, $f=0.786$), and
($\alpha=6.91\times10^{-6}$, $f=0.993$), respectively. In these
plots as before we have used the values $C_\gamma=70$,
$\rho_{Ch0}=1$, $A=0.775$, $\beta=0.00126$ and $m_p=1$ .
 \label{fig4}}}
\end{figure}

\begin{figure}[th]
{{\hspace{-1.4in}\includegraphics[width=4.2in,angle=0,clip=true]{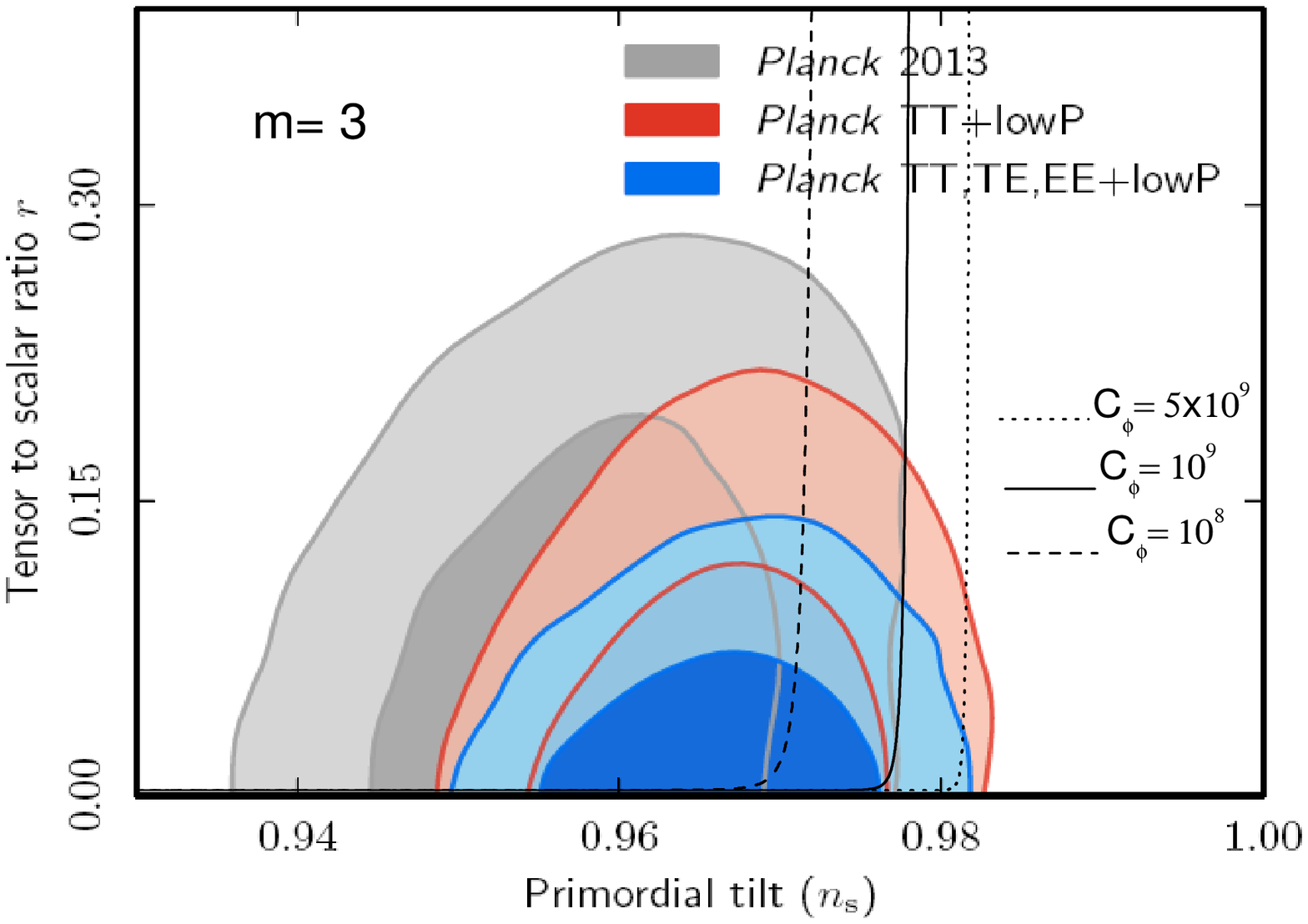}}}

{\vspace{-1.8 cm}\caption{ Evolution of the  tensor-to-scalar
ratio $r$ versus the scalar spectral index $n_s$ in the strong
dissipative regime for the special case $\Gamma\propto T^3/\phi^2$
(or analogously $m=3$). Here, we have considered the
two-dimensional marginalized constraints from the latest data of
Planck, see Ref.\cite{Planck2015}. Besides, we have studied three
different values of the parameter $C_\phi$. Also, in this plot the
dotted, solid,  and dashed lines correspond to the pairs
($\alpha=1.46\times10^{-5}$, $f=0.703$),
($\alpha=6.91\times10^{-6}$, $f=0.786$), and
($\alpha=6.91\times10^{-6}$, $f=0.993$), respectively. As before,
in this plot we have used the values $C_\gamma=70$,
$\rho_{Ch0}=1$, $A=0.775$, $\beta=0.00126$ and $m_p=1$.
 \label{fig5}}}
\end{figure}

In Fig.\ref{fig4} we show the evolution of the ratio $\Gamma/3H$
(left panel) and $T/H$ (right panel) on the scalar spectral index
$n_s$ in the strong dissipative regime, in the case in which the
dissipative coefficient $\Gamma\propto T^3/\phi^2$ (or analogously
$m=3$). In both panels we have studied three different values of
the parameter $C_\phi$. In order to write down the ratio
$\Gamma/3H$ versus $n_s$ (left panel), we have obtained
numerically, from Eqs.(\ref{nsf3N}) and (\ref{rs3N}), the relation
$\Gamma/3H=\Gamma/3H(n_s)$ for the case $m=3$. Likewise, this case
we have found numerically the evolution of the relation
$T/H=T/H(n_s)$ (right panel), considering Eqs.(\ref{rh-1}),
(\ref{Hf3}) and (\ref{Nsr3}) during the strong dissipative regime.
As before, in theses plots we have used the values $C_\gamma=70$,
$\rho_{Ch0}=1$, $A=0.775$, $\beta=0.00126$ \cite{const2} and
$m_p=1$. Analogously as in the case of the weak dissipative
regime,  we have found numerically from Eqs.(\ref{Prf3N}) and
(\ref{nsf3N}), for the special case $m=3$, the pair
($\alpha=1.46\times10^{-5}$, $f=0.703$), corresponding to the
parameter $C_\phi=5\times 10^{9}$, using the values
$\mathcal{P}_{\mathcal{R}}= 2.4\times 10^{-9}$, $n_s=0.97$ and the
number of $e-$folds $N=60$. Similarly, for the value of
$C_\phi=10^{9}$ we obtain numerically the pair
($\alpha=6.91\times10^{-6}$, $f=0.786$) and for parameter
$C_\phi=10^{8}$ we find ($\alpha=6.91\times10^{-6}$, $f=0.993$).
From the left panel we note that the values $C_\phi> 10^8$ satisfy
the condition for the strong dissipative regime. In this way, the
condition $\Gamma/3H>1$, gives a lower bound for the parameter
$C_\phi$. Also, we see that the essential condition for warm
inflation $T>H$, is well corroborated from the figure of the right
panel, and in fact, this condition does not impose a constraint on
the parameter $C_\phi$, in the strong dissipative regime.

In Fig.\ref{fig5} we show the dependence of the tensor-to-scalar
ratio on the scalar spectral index. Considering
Ref.\cite{Planck2015}, we have the two-dimensional marginalized
constraints at 68$\%$ and 95$\%$ confidence levels on the
parameters $r$ and $n_s$. As before, we have
considered Eqs.(\ref{nsf3N}) and (\ref{rs3N}) for the case $m=3$,
and we find numerically the consistency relation $r=r(n_s)$. Here
we observe that for the value of $C_\phi<5\times10^9$, the model
in the strong dissipative regime is well corroborated by the
observational data. In this way, for the case in
which the dissipative coefficient is given by $\Gamma\propto
T^3/\phi^2$ (or equivalently $m=3$), the range obtained for the parameter
$C_\phi$ is
$10^8<C_\phi<5\times10^9$.

Now considering the case in which $\Gamma\propto T$ (or
equivalently $m=1$) during the strong dissipative regime, here we
find from the condition $\Gamma>3H$, that the lower bound
for $C_\phi$ becomes $C_\phi>0.02$. In this form, the
condition for the strong
regime gives a lower bound on the parameter $C_\phi$ (figure not
shown). Similarly, considering the  consistency relation
$r=r(n_s)$ from the two-dimensional marginalized constraints
by the Planck data, we observe that these values are
well corroborated. Moreover, the tensor-to-scalar ratio becomes $r\sim 0$.
Similarly, for values of $C_\phi>0.02$, we observe that the
condition of warm inflation $T>H$ also is satisfied.  In this way,
we find only a lower bound for the parameter $C_\phi$ from
the condition $\Gamma/3H>1$. Then for the special case in which
$\Gamma\propto T$ (or equivalently $m=1$) the constraint for the
parameter $C_\phi$ results $C_\phi>0.02$.

For the cases in which $\Gamma\propto \phi$ (or equivalently
$m=0$) and $\Gamma\propto \phi^2/T$ (or equivalently $m=-1$), we
obtain that these models do not work in the strong
dissipative regime, since the scalar spectral index $n_s>1$, and then
these models are disproved from the observational data.
\\
\\
\begin{table}
\centering
\begin{tabular}
[c]{|c|c|c|}\hline $\Gamma=\frac{C_{\phi}T^{m}}{\phi^{m-1}}$ &
Constraints on $\alpha$ and $f$ & Constraint on $C_{\phi}$\\\hline
$m=3$ & $%
\begin{tabular}
[c]{c}%
$8.151\times10^{-7}<\alpha<1.461\times10^{-5}$\\
$0.703<f<0.993$%
\end{tabular}
\ \ $ & $10^{8}<C_{\phi}<5\times10^{9}$\\\hline $m=1$ &
\begin{tabular}
[c]{c}%
$\alpha<4.539$\\
$f>0.204$%
\end{tabular}
& $C_{\phi}>0.02$\\\hline $m=0$ & The model does not work
$(n_{s}>1)$ & --\\\hline $m=-1$ & The model does not work
$(n_{s}>1)$ & --\\\hline
\end{tabular}
\caption{Results for the constraints on the parameters $\alpha$,
$f$ and $C_\phi$ during the strong dissipative regime.} \label{T2}
\end{table}

Table \ref{T2} indicates the constraints on the parameters
$\alpha$, $f$ and $C_\phi$, for  different values of the parameter
$m$, considering a general  form for the dissipative coefficient
$\Gamma=\Gamma(T,\phi)$, in the strong dissipative regime. We
observe that for the special case $m=3$,  the constraints on our
parameters, result as consequence of the condition $\Gamma>3H$
(lower bound), and from the consistency  relation $r=r(n_s)$(upper
bound). For the case $m=1$, we find only a lower bound from the
condition $\Gamma>3H$.
Here we have used  the values
$C_\gamma=70$, $\rho_{Ch0}=1$, $A=0.775$, $\beta=0.00126$ and
$m_p=1$.

\begin{figure}[th]
{{{\vspace{0cm}\includegraphics[width=2.6in,angle=0,clip=true]{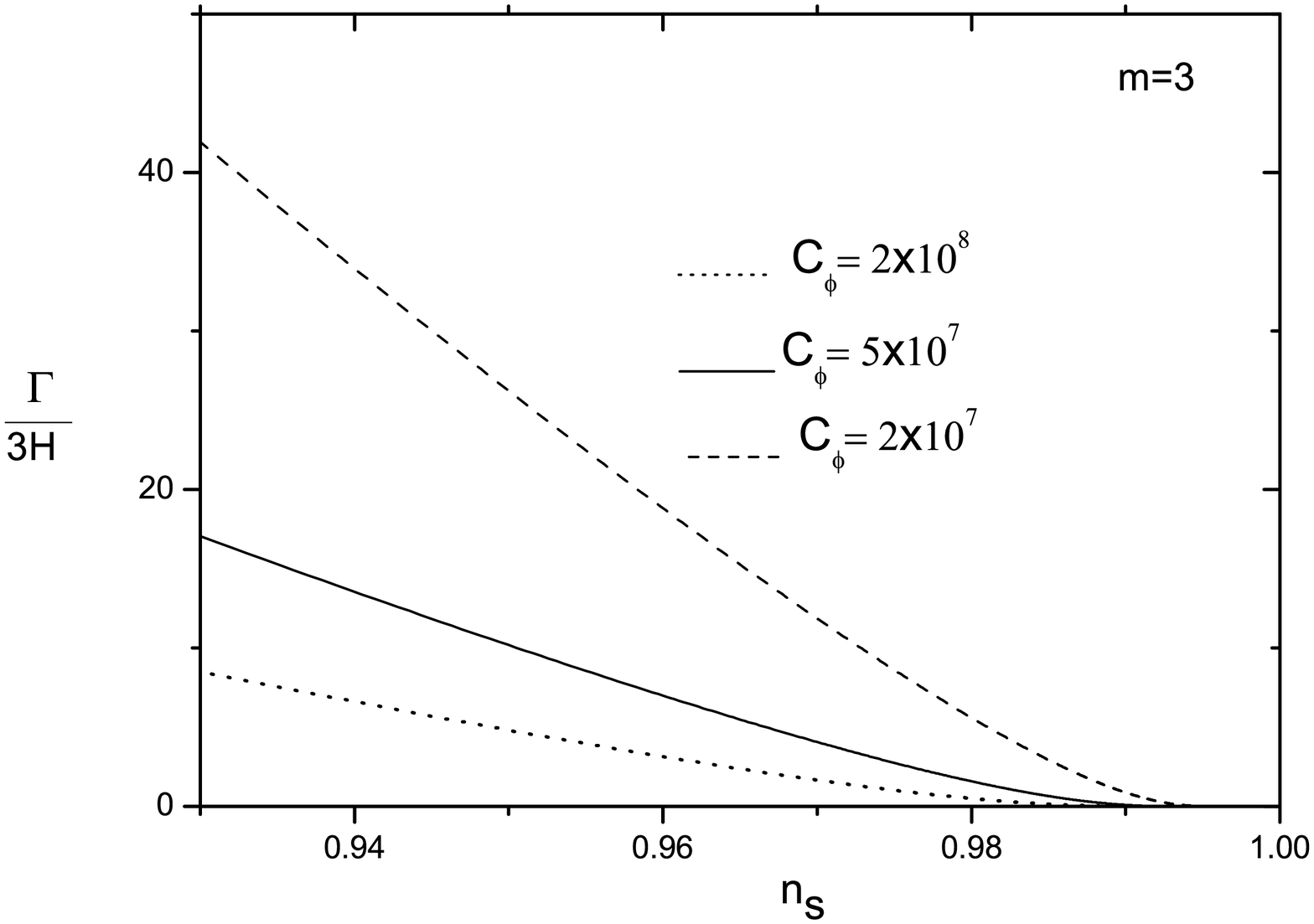}}}}\\
{\hspace{-3.3 cm}{{\vspace{-1.
cm}\includegraphics[width=3.5in,angle=0,clip=true]{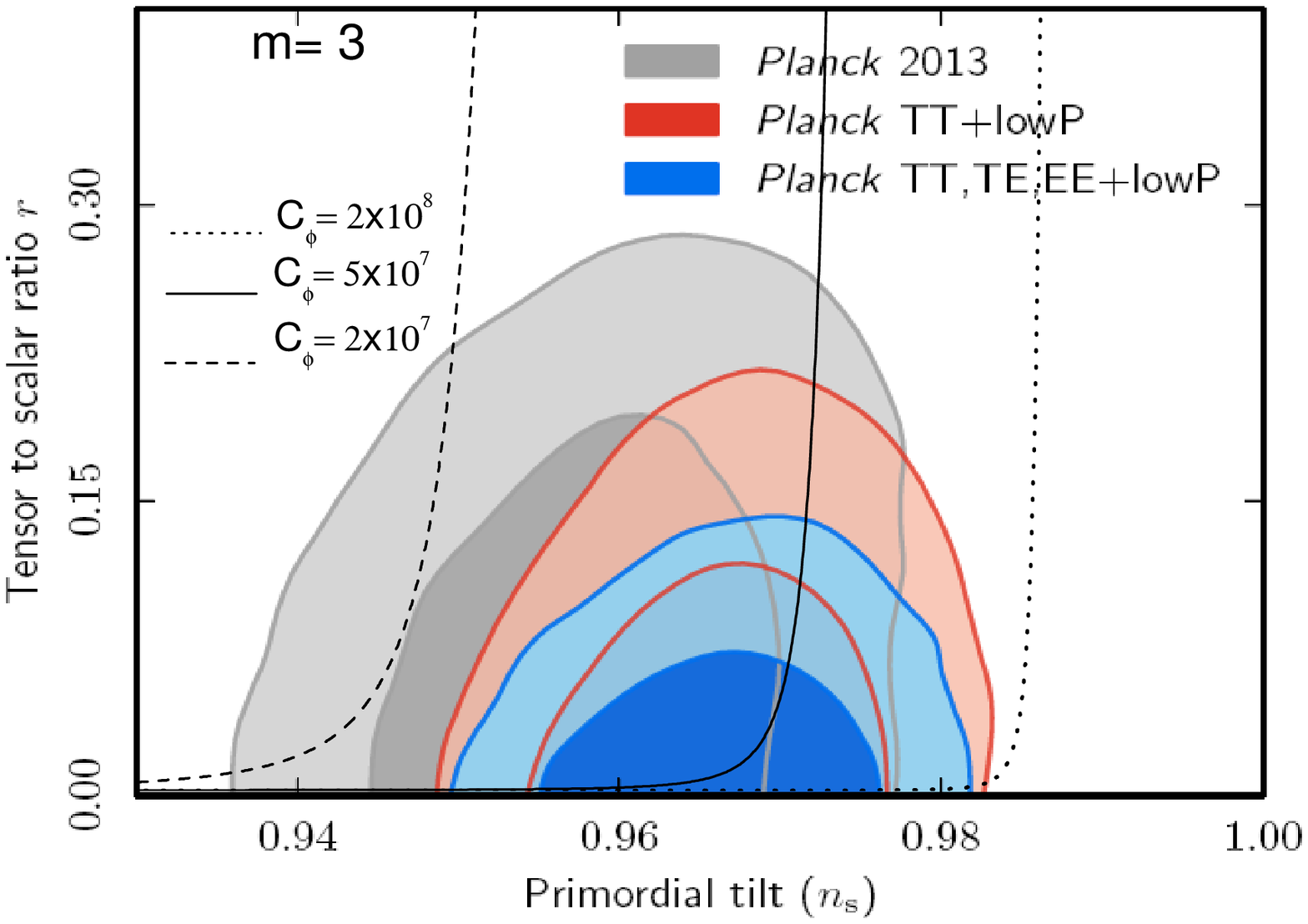}}}}
{\vspace{-1.0 cm}\caption{ Upper panel:  ratio $\Gamma/3H$ versus
the scalar spectral index $n_s$. Lower panel: ratio $r$ versus the
scalar spectral index $n_s$. For both panels we have used
different values of the parameter $C_\phi$, for the special case
$m=3$ or equivalently $\Gamma\propto T^3/\phi^2$ during  the
strong dissipative regime. In both panels, the dotted, solid, and
dashed lines correspond to the pairs ($A=0.22$, $\beta=-0.96$),
($A=0.06$, $\beta=-0.97$), and ($A=0.04$, $\beta=-0.98$),
respectively. In these plots  we have used the values
$C_\gamma=70$, $\rho_{Ch0}=1$, $\alpha=10^{-3}$, $f=0.6$ and
$m_p=1$ .
 \label{fig6c}}}
\end{figure}

 Analogously to the case of the weak dissipative regime, we
can also numerically obtain results for the parameters $A$ and $\beta$,
from the observational constraints for the power spectrum and the scalar spectral index at $N=60$. In this way, we can fix the
values $\alpha$, $f$ and $C_\phi$.  In Fig.(\ref{fig6c}) we show
the plot of the ratio $\Gamma/3H$ (upper panel) and the
tensor-to-scalar ratio $r$ (lower panel) as functions of the primordial tilt $n_s$
for the specific case $m=3$, in the strong
dissipative regime. As before, for both panels we have considered three values for $C_\phi$. In the upper plot we show the decay of the ratio
 $R=\Gamma/3H$ during inflation, however always satisfying the condition $\Gamma>3H$, in agreement with strong dissipative regime. In the lower panel we show the
 two-dimensional constraints on the parameters $r$ and $n_s$ from Planck 2015 data.

As before, we numerically find the ratio $\Gamma/3H$, and the tensor-to-scalar
ratio as functions of the scalar spectral index $n_s$,
considering  Eqs.(\ref{gamma3}), (\ref{rs3N}) and (\ref{nsf3N}).
Now, we use the values $C_\gamma=70$,
$\rho_{Ch0}=1$, $\alpha=10^{-3}$, and  $f=0.6$. For the special case $m=3$, we
obtain numerically that the
values $A=0.22$ and $\beta=-0.98$ correspond to the
parameter $C_\phi=2\times 10^{8}$. As usual, we have considered  the values
${\mathcal{P}_{\mathcal{R}}}=2.43\times10^{-9}$,
$n_s=0.97$, and the number of $e$-folds $N=60$.  Similarly, for the value of $C_\phi=5\times10^{8}$, we
obtain
the values  $A=0.062$ and $\beta=-0.97$. Finally,
for the parameter $C_\phi=2 \times10^{8}$, we obtain the values $A=0.04$ and $\beta=-0.98$.
From the upper plot, we may
obtain  an upper limit for the parameter $C_\phi$, given by $C_\phi<2\times10^8$
which satisfies $\Gamma>3H$. Analogously,  from the upper plot
we obtain  a lower limit, given by $C_\phi>2\times10^7$,  from the
consistency relation $r=r(n_s)$.  On the other hand, from the essential condition for
warm inflation to occur, i.e., $T>H$, we note that this condition does not
impose any constraint on the parameter $C_\phi$, since the condition $T>H$ is always satisfied (plot not shown). Therefore,
for the value $m=3$, the allowed range for $C_{\phi}$ is given by $2\times10^7<C_\phi<2\times10^8$.

For the case  $m=1$,  and considering  the condition for the
strong dissipative regime $\Gamma>3H$, we obtain a lower bound
for the parameter $C_\phi$. Analogously as the case $m=3$, we
fixed the values $\alpha=10^{-3}$ and $f=0.6$. In this way, we
numerically obtain that the values $A=0.45$ and $\beta=-4.2$,
correspond to $C_\phi=2\times10^{-2}$, for which
$\Gamma/3H\gtrsim 1$ (not shown).For the specific case $m=1$, we observe that this value for $C_\phi$ is allowed by the
Planck 2015 data. For this value, the tensor-to-scalar ratio becomes $r< 0.05$ and, by the other hand, the essential condition for
warm inflation $T>H$ is always satisfied(not shown).
In this way, for $m=1$
we only obtain a lower bound for the parameter $C_\phi$, from
the condition $\Gamma/3H>1$. Then for the special case in which
$\Gamma\propto T$ (or equivalently $m=1$) the constraints on the
parameters are given by  $C_\phi>2\times10^{-2}$, $\beta>-4.2$ and $0<A<0.45$.

For the cases $\Gamma\propto \phi$ (
$m=0$) and $\Gamma\propto \phi^2/T$ ($m=-1$), we
find that these models do not work in the strong
dissipative regime, because the scalar spectral index $n_s>1$.
\\
\\
\begin{table}
\centering
\begin{tabular}
[c]{|c|c|c|}\hline $\Gamma=\frac{C_{\phi}T^{m}}{\phi^{m-1}}$ &
Constraints on $A$ and $\beta$ & Constraint on $C_{\phi}$\\\hline
$m=3$ & $%
\begin{tabular}
[c]{c}%
$0.04<A<0.22$\\
$-0.98<\beta<-0.96$%
\end{tabular}
\ \ $ & $10^{8}<C_{\phi}<5\times10^{9}$\\\hline $m=1$ &
\begin{tabular}
[c]{c}%
$0<A<0.45$\\
$\beta>-4.2$%
\end{tabular}
& $C_{\phi}>2\times10^{-2}$\\\hline $m=0$ & The model does not work
$(n_{s}>1)$ & --\\\hline $m=-1$ & The model does not work
$(n_{s}>1)$ & --\\\hline
\end{tabular}
\caption{Results for the constraints on the parameters $A$,
$\beta$ and $C_\phi$ during the strong dissipative regime.} \label{T4}
\end{table}

Table \ref{T4} shows the constraints on the parameters
$A$, $\beta$ and $C_\phi$, for  different values of the parameter
$m$ in the strong dissipative regime. We
observe that for the special case $m=3$,  the constraints on these
parameters result as consequence of the condition $\Gamma>3H$
(lower bound), and from the consistency  relation $r=r(n_s)$(upper
bound). For the case $m=1$, we only find a lower bound from the
condition $\Gamma>3H$, and for the cases  $m=0$ and $m=-1$, these models do not work.
 Here we have used  the values
$C_\gamma=70$, $\rho_{Ch0}=1$, $\alpha=10^{-3}$, $f=0.6$ and
$m_p=1$.

\section{Conclusions \label{conclu}}

In this paper we have studied the warm-intermediate inflationary
model in the context of Generalized Chaplygin Gas as a variant of gravity. For the weak and strong
dissipative regimes,
we have  found solutions to the background equations under the
slow-roll approximation. Here, we have considered a general form
of the dissipative coefficient $\Gamma\propto\,T^{m}/\phi^{m-1}$,
and we have analyzed the cases $m=3$, $m=1$, $m=0$ and $m=-1$.
By the other hand, we have obtained expressions for  the scalar and tensor power spectrum,
scalar spectral index, and tensor-to-scalar ratio. For both
regimes, we have found the constraints on the several parameters,
considering the Planck 2015 data, together with the condition for
warm inflation $T>H$, and the condition for the weak $\Gamma<3H$ (or
strong $\Gamma>3H$) dissipative regime.

In our analysis for both regimes, in first place we have fixed  the parameters $A$ and $\beta$, and then
we have found different constraints on the parameters $\alpha$, $f$ and $C_\phi$. Secondly,
we have fixed the parameters $A$ and $f$, and then
we have obtained  constraints over the parameters $A$, $\beta$ and $C_\phi$.  In this latter case,
we have found that negative values for $\beta$ are allowed, and also the results weakly depend on the parameter $\beta$ in both regimes.

For the weak dissipative regime we have obtained the constraints
on the parameters of our model, only from the conditions
$\Gamma<3H$, which gives an upper bound, and $T>H$, which gives a
lower bound. This is due that fact that the consistency relation
$r=r(n_s)$ does not impose  constraints on the parameters. For the
strong dissipative regime, we have found the constraints on the
parameters from the Planck 2015 data, trough the consistency
relation $r=r(n_s)$, and the condition $\Gamma>3H$. Here, the
condition for warm inflation $T>H$ does not give constraints on
the parameters. By the other hand, for the strong dissipative
regime we have obtained that for the cases  $\Gamma\propto \phi $
( or equivalently  $m=0$) and $\Gamma\propto \phi^2/T$ ( or
equivalently $m=-1$) these models do not work, since the scalar
spectral index becomes $n_s>1$, so predicting a blue tilted spectrum, and then
these models are disproved from observational data.

 On the other hand, we have observed that when the value of the parameter $m$
decreases, the values belonging to the allowed range for
$C_{\phi}$ also decrease.

Summarizing, only the cases $m=3$ ($\Gamma\propto T^3/\phi^2$) and $m=1$
($\Gamma\propto T$) of the generalized dissipative coefficient, given by Eq.(\ref{G}), describe successfully  a
warm-intermediate inflationary model in the context of GCG.
These models are well supported by the Planck 2015 data, through
the consistency relation $r=r(n_s)$, and satisfy
 the essential condition for warm inflation $T>H$, and the requirement to
evolve according to the weak ($\Gamma<3H$), or strong ($\Gamma>3H$) dissipative regime. Our results are
summarized in Tables I, II, III and VI, respectively.

\begin{acknowledgments}
R.H. was supported by Comisi\'on Nacional de Ciencias y
Tecnolog\'ia of Chile through FONDECYT Grant N$^{0}$ 1130628 and
DI-PUCV N$^{0}$ 123.724. N.V. was supported by Comisi\'on Nacional
de Ciencias y Tecnolog\'ia of Chile through FONDECYT Grant N$^{0}$
3150490.
\end{acknowledgments}

%\\\\\\\\\\\\\\\\\\\\\\\\\\\\\\\\\\\\\\\\\\\\\\\\\\\\\\\\\\\\\\\\\\\\\\\


\begin{thebibliography}{99}                                                                                               %



%%%%%Refer introduction%%%%%%%%%%%%%%%%%%%%%%%%%%%%%%%%%%


\bibitem {R1}A. Guth , Phys. Rev. D \textbf{23}, 347 (1981).

\bibitem {R102}A.A. Starobinsky, Phys. Lett. B \textbf{91}, 99 (1980).

\bibitem {R103}A.D. Linde, Phys. Lett. B \textbf{108}, 389 (1982).

\bibitem {R104}A.D. Linde, Phys. Lett. B \textbf{129}, 177 (1983).

\bibitem {R105}A. Albrecht and P. J. Steinhardt, Phys. Rev. Lett.
\textbf{48},1220 (1982).

\bibitem {R106}K. Sato, Mon. Not. Roy. Astron. Soc. \textbf{195}, 467 (1981).

\bibitem {R2}V.F. Mukhanov and G.V. Chibisov , JETP Letters \textbf{33},
532(1981).

\bibitem {R202}S. W. Hawking,Phys. Lett. B \textbf{115}, 295 (1982).

\bibitem {R203}A. Guth and S.-Y. Pi, Phys. Rev. Lett. \textbf{49}, 1110 (1982).

\bibitem {R204}A. A. Starobinsky, Phys. Lett. B \textbf{117}, 175 (1982).

\bibitem {R205}J.M. Bardeen, P.J. Steinhardt and M.S. Turner, Phys. Rev.D
\textbf{28}, 679 (1983).

\bibitem {astro}D.~Larson \textit{et al.},
%``Seven-Year Wilkinson Microwave Anisotropy Probe (WMAP) Observations: Power
%Spectra and WMAP-Derived Parameters,''
Astrophys.\ J.\ Suppl.\ \textbf{192}, 16 (2011).

\bibitem {astro2}C.~L.~Bennett \textit{et al.},
%``Seven-Year Wilkinson Microwave Anisotropy Probe (WMAP) Observations: Are
%There Cosmic Microwave Background Anomalies?,''
Astrophys.\ J.\ Suppl.\ \textbf{192}, 17 (2011).

\bibitem {astro202}N.~Jarosik \textit{et al.},
%``Seven-Year Wilkinson Microwave Anisotropy Probe (WMAP) Observations: Sky
%Maps, Systematic Errors, and Basic Results,''
Astrophys.\ J.\ Suppl.\ \textbf{192}, 14 (2011).



\bibitem{Planck2015}
  P.~A.~R.~Ade {\it et al.}  [Planck Collaboration],
  %``Planck 2015 results. XX. Constraints on inflation,''
  arXiv:1502.02114 [astro-ph.CO].

\bibitem {warm}A. Berera, Phys. Rev. Lett. \textbf{75}, 3218 (1995); A.
Berera, Phys. Rev. D \textbf{55}, 3346 (1997).

\bibitem {62526}L.M.H. Hall, I.G. Moss and A. Berera, Phys.Rev.D \textbf{69},
083525 (2004)
\bibitem {1126}A. Berera, Phys. Rev.D \textbf{54}, 2519 (1996).
\bibitem {Berera:2008ar}A.~Berera, I.~G.~Moss and R.~O.~Ramos,
%``Warm Inflation and its Microphysical Basis,''
Rept.\ Prog.\ Phys.\ \textbf{72}, 026901 (2009); M.~Bastero-Gil, A.~Berera, I.~G.~Moss and R.~O.~Ramos,
  %``Theory of non-Gaussianity in warm inflation,''
  JCAP {\bf 1412}, no. 12, 008 (2014); S.~Bartrum, M.~Bastero-Gil, A.~Berera, R.~Cerezo, R.~O.~Ramos and J.~G.~Rosa,
  %``The importance of being warm (during inflation),''
  Phys.\ Lett.\ B {\bf 732}, 116 (2014).

\bibitem{Chap}A. Kamenshchik, U. Moschella and V. Pasquier, Phys. Lett. B 511, 265
(2001).

%\bibitem{const}U. Alam, V. Sahni, T.D. Saini, and A.A. Starobinsky,
%Mon. Not. R. Astron. Soc. 344, 1057 (2003); L. Amendola, F.
%Finelli, C. Burigana, and D. Carturan, JCAP 0307, 005 (2003); X.
%Zhang, F.-Q. Wu, and J. Zhang, JCAP 0601, 003 (2006); L. Xu, J.
%Lu, JCAP 1003, 025(2010); J. Lu, Y. Gui, L. Xu, Eur. Phys. J. C
%63,349(2009); Z. Li, P. Wu, H. Yu, JCAP09,017(2009).



\bibitem{F2}J.~C.~Fabris, T.~C.~C.~Guio, M.~Hamani Daouda and O.~F.~Piattella,
  %``Scalar models for the generalized Chaplygin gas and the structure formation constraints,''
  Grav.\ Cosmol.\  {\bf 17}, 259 (2011).
\bibitem{F1}R. Colistete Jr, J. C. Fabris, S.V.B. Gongalves and P.E. de
Souza, Int. J. Mod. Phys. D{\bf 13}, 669(2004); R. Colistete Jr.
and J. C. Fabris, Class. Quant. Grav. {\bf22}, 2813 (2005).

















\bibitem{const1}   N.
Liang, L. Xu, Z. H. Zhu, Astrono.  Astrophy, 527, A11(2011); C. G.
Park, J. c. Hwang, J. Park, H. Noh, Phys. Rev. D 81,063532(2010).

\bibitem{271} M. C. Bento, O. Bertolami and A. Sen, Phys. Rev. D 66, 043507
(2002).
\bibitem{const2}L.~Xu, J.~Lu, Y.~Wang, J.~Lu and Y.~Wang,
  %``Revisiting Generalized Chaplygin Gas as a Unified Dark Matter and Dark Energy Model,''
  Eur.\ Phys.\ J.\ C {\bf 72}, 1883 (2012).

\bibitem{Avelino:2015dwa} S.~del Campo, C.~R.~Fadragas, R.~Herrera, C.~Leiva, G.~Leon and J.~Saavedra,
  %``Thawing models in the presence of a generalized Chaplygin gas,''
  Phys.\ Rev.\ D {\bf 88}, 023532 (2013); R.~Herrera, M.~Olivares and N.~Videla,
  %``Intermediate-Generalized Chaplygin Gas inflationary universe model,''
  Eur.\ Phys.\ J.\ C {\bf 73}, no. 1, 2295 (2013);
  P.~P.~Avelino and V.~M.~C.~Ferreira,
  %``Constraints on the dark matter sound speed from galactic scales: the cases of the Modified and Extended Chaplygin Gas,''
  Phys.\ Rev.\ D {\bf 91}, no. 8, 083508 (2015).





\bibitem{Bertolami:2006zg}
  O.~Bertolami and V.~Duvvuri,
  %``Chaplygin inflation,''
  Phys.\ Lett.\ B {\bf 640}, 121 (2006).



\bibitem{mon1} S. del Campo and R. Herrera, Phys. Lett. B 660, 282
(2008); R.~Herrera,
  %``Tachyon-Chaplygin inflation on the brane,''
  Gen.\ Rel.\ Grav.\  {\bf 41}, 1259 (2009);R.~Zarrouki and M.~Bennai,
  %``Chaplygin gas braneworld inflation according to WMAP7 data,''
   Phys.\ Rev.\ D {\bf 82}, 123506 (2010).
\bibitem{mod}
L. Randall and R. Sundrum, Phys. Rev. Lett. 83, 4690 (1999); T.
Shiromizu, K. Maeda and M. Sasaki, Phys. Rev. D 62, 024012 (2000);
R. Maartens, Lect. Notes Phys. 653 213 (2004); A. Lue, Phys. Rept.
423, 1 (2006).
\bibitem{mod2}T.~Clifton, P.~G.~Ferreira, A.~Padilla and C.~Skordis,
  %``Modified Gravity and Cosmology,''
  Phys.\ Rept.\  {\bf 513}, 1 (2012).
\bibitem{power}F. Lucchin and S. Matarrese, Phys. Rev. D32, 1316 (1985).

\bibitem{Barrow1} J. D Barrow,
Phys. Lett. B {\bf 235}, 40 (1990); J. D Barrow and P. Saich,
Phys. Lett. B {\bf 249}, 406 (1990);A. Muslimov, Class. Quantum
Grav. {\bf 7}, 231 (1990); A. D. Rendall, Class. Quantum Grav.
{\bf 22}, 1655 (2005).

\bibitem{Barrow2} J. D Barrow and A. R. Liddle,
Phys. Rev. D {\bf 47}, R5219 (1993);  A. A. Starobinsky JETP Lett.
{\bf 82}, 169 (2005); S.~del Campo, R.~Herrera, J.~Saavedra,
C.~Campuzano and E.~Rojas,
  %``Curvaton reheating in logamediate inflationary model,''
  Phys.\ Rev.\  D {\bf 80}, 123531 (2009);   R.~Herrera and E.~San Martin,
  %``Warm-intermediate inflationary universe model in braneworld cosmologies,''
  Eur.\ Phys.\ J.\ C {\bf 71}, 1701 (2011); R.~Herrera and M.~Olivares,
  %``Logamediate inflation on the brane,''
  Mod.\ Phys.\ Lett.\ A {\bf 27}, 1250101 (2012); R.~Herrera and M.~Olivares,
  %``Warm-Logamediate inflationary universe model,''
  Int.\ J.\ Mod.\ Phys.\ D {\bf 21}, 1250047 (2012); R.~O.~Ramos and L.~A.~da Silva,
  %``Power spectrum for inflation models with quantum and thermal noises,''
  JCAP {\bf 1303}, 032 (2013).

\bibitem{ratior}
W. H. Kinney, E. W. Kolb, A. Melchiorri and A. Riotto, Phys. Rev.
D {\bf 74}, 023502 (2006); R. Herrera and E. San Martin,  Int.\
J.\ Mod.\ Phys.\ D {\bf 22}, 1350008 (2013).

\bibitem{Barrow3} J. D. Barrow, A. R. Liddle and C. Pahud, Phys. Rev. D, {\bf 74}, 127305
(2006); R.~Herrera,
%``Warm inflationary model in loop quantum cosmology,''
Phys.\ Rev.\ \textbf{D81}, 123511 (2010).


\bibitem{KM}T. Kolvisto and D. Mota, Phys. Lett. {\bf B 644},
104 (2007); Phys. Rev. D. {\bf 75}, 023518 (2007).

\bibitem{ART}I. Antoniadis, J. Rizos and K. Tamvakis, Nucl.Phys.
{\bf B 415}, 497 (1994).

\bibitem{BD}D. G. Boulware and S. Deser, Phys.Rev. Lett.
{\bf 55}, 2656 (1985); Phys. Lett. {\bf B 175}, 409 (1986).


\bibitem{Varios1}S. Mignemi and N. R. Steward, Phys. Rev. D {\bf 47}, 5259 (1993);
P. Kanti, N. E. Mavromatos, J. Rizos, K. Tamvakis and E.
Winstanley, Phys. Rev. D {\bf 54}, 5049 (1996); Ch.M Chen, D. V.
Gal'tsov and D. G. Orlov, Phys. Rev. D {\bf 75}, 084030 (2007).

\bibitem{Varios2}S. Nojiri, S. D. Odintsov and M. Sasaki, Phys. Rev. D {\bf 71},
123509 (2004); G. Gognola, E. Eizalde, S. Nojiri, S. D. Odintsov
and E. Winstanley, Phys. Rev. D {\bf 73}, 084007 (2006).

\bibitem{Sanyal}A. K. Sanyal, Phys. Lett. {\bf B}, {\bf 645},1
(2007).







\bibitem{new1}I. Antoniadis, J. Rizos and K. Tamvakis, Nucl. Phys. B.
415, 497 (1994); S. Nojiri, S. D. Odintsov and M. Sasaki, Phys.
Rev. D. 71, 123509 (2004); G. Gognola, E. Eizalde, S. Nojiri, S.
D. Odintsov and E. Winstanley, Phys. Rev. D. 73, 084007 (2006).


\bibitem{varios1} A.~Cid, G.~Leon and Y.~Leyva,
  %``Intermediate accelerated solutions as generic late-time attractors in a modified Jordan-Brans-Dicke theory,''
  arXiv:1506.00186 [gr-qc]; R.~Herrera, N.~Videla and M.~Olivares,
  %``Warm intermediate inflation in the Randall–Sundrum II model in the light of Planck 2015 and BICEP2 results: a general dissipative coefficient,''
  Eur.\ Phys.\ J.\ C {\bf 75}, no. 5, 205 (2015); R.~Herrera, S.~del Campo, M.~Olivares, J.~Saavedra and N.~Videla,
  %``Intermediate inflation on warped DGP model,''
  AIP Conf.\ Proc.\  {\bf 1647}, 104 (2015); R.~Herrera, N.~Videla and M.~Olivares,
  %``Warped DGP model in warm intermediate inflation with a general dissipative coefficient in light of BICEP2 and Planck results,''
  Phys.\ Rev.\ D {\bf 90}, no. 10, 103502 (2014); R.~Herrera, M.~Olivares and N.~Videla,
  %``General dissipative coefficient in warm intermediate inflation in loop quantum cosmology in light of Planck and BICEP2,''
  Int.\ J.\ Mod.\ Phys.\ D {\bf 23}, no. 10, 1450080 (2014);
   J.~D.~Barrow and J.~Magueijo,
  %``Intermediate inflation from rainbow gravity,''
  Phys.\ Rev.\ D {\bf 88}, no. 10, 103525 (2013).


\bibitem{gamma1}Y. Zhang, JCAP 0903, 023 (2009);
 M. Bastero-Gil, A. Berera and R. O. Ramos, JCAP 1107, 030 (2011).

\bibitem {new2}G. Calcagni and G. Nardelli, Nucl. Phys. B \textbf{823}, 234
(2009).

%%%%%%%%%%%%%%%%%%%%%%%%%%%%%%%%%%%%%%%%%%%%%%%%%%%%%%%%%%%%%%%%%%%%%%%%%%%%%%%%%


\bibitem {26}I.~G.~Moss and C.~Xiong,
%``Dissipation coefficients for supersymmetric inflatonary models,''
arXiv:hep-ph/0603266.

\bibitem {28}A.~Berera, M.~Gleiser and R.~O.~Ramos,
%``Strong dissipative behavior in quantum field theory,''
Phys.\ Rev.\ D \textbf{58} 123508 (1998).

\bibitem {2802}A.~Berera and R.~O.~Ramos,
%``The affinity for scalar fields to dissipate,''
Phys.\ Rev.\ D \textbf{63}, 103509 (2001).


\bibitem {6252602}I.G. Moss, Phys.Lett.B \textbf{154}, 120 (1985).

\bibitem {6252603}A.Berera and L.Z. Fang, Phys.Rev.Lett. \textbf{74} 1912
(1995).

\bibitem {6252604}A.Berera, Nucl.Phys B \textbf{585}, 666 (2000).

\bibitem {Libro}Abramowitz, M. and Stegun, I. A. (Eds.). Handbook of
Mathematical Functions with Formulas, Graphs, and Mathematical Tables, 9th
printing. New York: Dover, 1972.

\bibitem {B1}A. Berera, Nucl. Phys. B \textbf{585}, 666 (2000).

\bibitem{Taylor:2000ze}
  A.~N.~Taylor and A.~Berera,
  %``Perturbation spectra in the warm inflationary scenario,''
  Phys.\ Rev.\ D {\bf 62}, 083517 (2000)

%%%%%%%%%%%%%%%%%%%%%%%%%%%%%%%%%%%%%%%%%%%%%%%%%%%%%%

\bibitem{I1} L. Z. Fang, Phys. Lett. B {\bf95}, 154 (1980).


\bibitem{I2} I. G. Moss, Phys. Lett. B {\bf154}, 120 (1985);
 J. Yokoyama and K. I. Maeda, Phys. Lett. B {\bf207}, 31 (1988).







%%%%%%%%%%%%%%%%%%%%%%%%%%%%%

%\bibitem{NG0} P.~A.~R.~Ade {\it et al.} [Planck Collaboration],
  %``Planck 2015 results. XVII. Constraints on primordial non-Gaussianity,''
  %arXiv:1502.01592 [astro-ph.CO].

























%\bibitem {taylorberera}A. Berera, Phys. Rev. D \textbf{55}, 3346 (1997)

%\bibitem {taylorberera02}J. Mimoso, A. Nunes and D. Pavon, Phys.Rev.D
%\textbf{73}, 023502 (2006)

%\bibitem {taylorberera03}R.~Herrera, S.~del Campo and C.~Campuzano,
%``Tachyon warm inflationary universe models,''
%JCAP \textbf{10}, 009 (2006)

%\bibitem {taylorberera04}S. del Campo, R. Herrera and D. Pavon, Phys. Rev. D
%\textbf{75}, 083518 (2007)

%\bibitem {taylorberera05}S.~del Campo and R.~Herrera,
%``Warm inflation in the DGP brane-world model,''
%Phys.\ Lett.\ B \textbf{653}, 122 (2007)

%\bibitem {taylorberera06}M.~A.~Cid, S.~del Campo and R.~Herrera, JCAP
%\textbf{10}, 005 (2006)

%\bibitem {taylorberera07}J.~C.~B.~Sanchez, M.~Bastero-Gil, A.~Berera and
%K.~Dimopoulos, Phys.\ Rev.\ D \textbf{77} 123527 (2008)



%\bibitem {taylorberera09}R.~Herrera, E.~San Martin,
%``Warm-intermediate inflationary universe model in braneworld cosmologies,''
%Eur.\ Phys.\ J.\ \textbf{C71}, 1701 (2011).

%%%%%%%%%%%%%%%%%%%%%%%%%%%%%%%%%%%%%%%%%%%%%%%%


%%%%%%%%%%%%%%%%%%%INTERMEDIA


%%%%%%%%%%%%%%%%%%%%%%%%%%%%%%%%%%%%%%%%%%%%%%%%%%%%%%%%%%%%%%%%%%%%%%%%%%%%%%%%


%\bibitem {Bra}R. Brandenberger and M. Yamaguchi, Phys. Rev. D \textbf{68},
%023505 (2003).

%\bibitem {BasteroGil:2011cx}M.~Bastero-Gil, A.~Berera, R.~O.~Ramos and
%J.~G.~Rosa,
%``Warm baryogenesis,''
%arXiv:1110.3971 [hep-ph].

%\bibitem {R11}J. D. Barrow, Class. Quant. Grav. \textbf{13}, 2965 (1996).

%\bibitem {atp}J.~Yokoyama and K.~Maeda,
%``On the Dynamics of the Power Law Inflation Due to an Exponential
%Potential,''
%Phys.\ Lett.\ B \textbf{207}, 31 (1988).

%\bibitem {R12}J. D. Barrow and N. J. Nunes, Phys. Rev. D \textbf{76} 043501 (2007).

%\bibitem{R13}D.H. Lyth, A. Riotto, Phys. Rep. B {\bf314}, 1 (1999).


%\bibitem {R9}D. H. Lyth and A. Riotto, Phys. Rept. \textbf{314}, 1 (1999).

%\bibitem {R10}A. D. Linde (2005), hep-th/0503203.

%\bibitem {P1}P.G. Ferreira, M. Joyce, Phys.Rev.D 58, 023503 (1998).

%\bibitem {P2}P. Binetruy, Phys.Rev.D 60, 063502 (1999).

%\bibitem {A2}P. Parsons and J. D. Barrow, Phys. Rev. D \textbf{51}, 6757 (1995).

%\bibitem{R4}G. Smoot,  et al. Astrophys. J. Lett. {\bf396}, L1 (1992).


%\bibitem{R5} E.~Komatsu {\it et al.}  [WMAP Collaboration],
%``Five-Year Wilkinson Microwave Anisotropy Probe (WMAP\altaffilmark 1 )
%Observations:Cosmological Interpretation,''
%Astrophys.\ J.\ Suppl.\  {\bf 180}, 330 (2009).


%\bibitem{R6} L.P. Grishchuk,  Sov. Phys. JETP {\bf40}, 409 (1975); A.A. Starobinsky,
%JETP Lett. {\bf30}, 682 (1979).


%\bibitem{R7}G. Efstathiou, C. Lawrence  and J. Tauber  (coordinators), ESASCI(
%2005) 1.


%\bibitem{R8} http://lisa.jpl.nasa.gov/


%\bibitem {P3}Ph. Brax, J. Martin, Phys.Lett.B 468, 40 (1999).

%\bibitem {Ramon}S.~del Campo, R.~Herrera, J.~Saavedra, C.~Campuzano,
%E.~Rojas,
%``Curvaton reheating in logamediate inflationary model,''
%Phys.\ Rev.\ \textbf{D80}, 123531 (2009).

%\bibitem{P4} T. Damour, F. Piazza, G. Veneziano, Phys.Rev.D. 66, 046007
%(2002).
%\bibitem{P5} Ph. Brax, J. Martin, Phys.Rev.D 61, 103502 (2000);
%S.C.C. Ng, N.J. Nunes, F. Rosati, Phys.Rev.D 64, 083510
%(2001).


%\bibitem {P6}P.J.E. Peebles, B. Ratra, Rev.Mod.Phys.75, 559 (2003)

%\bibitem {P602}R.~Ghosh, S.~Chattopadhyay, U.~Debnath,
%``A Dark Energy Model with Generalized Uncertainty Principle in the Emergent, Intermediate and Logamediate Scenarios of the Universe,''
%[arXiv:1105.4538 [gr-qc]]

%\bibitem {P603}U.~Debnath, S.~Chattopadhyay, M.~Jamil,
%``Fractional Action Cosmology: Some Dark Energy Models in Emergent, Logamediate and Intermediate Scenarios of the Universe,''
%[arXiv:1107.0541 [physics.gen-ph]].

%\bibitem {new}G.~Calcagni and G.~Nardelli,
%``Cosmological rolling solutions of nonlocal theories,''
%Int.\ J.\ Mod.\ Phys.\ D \textbf{19}, 329 (2010).



%\bibitem {27}J.~C.~Bueno Sanchez, M.~Bastero-Gil, A.~Berera and
%K.~Dimopoulos,
%``Warm hilltop inflation,''
%Phys.\ Rev.\ D \textbf{77}, 123527 (2008).

%%%%%%%%%%%%%%%%




%\bibitem {BasteroGil:2010pb}M.~Bastero-Gil, A.~Berera and R.~O.~Ramos,
%``Dissipation coefficients from scalar and fermion quantum field
%interactions,''
%JCAP \textbf{1109}, 033 (2011).

%%%%%%%%%%%%%%%%%%%%%%%%%%%%%%%%%%%%%%%




%%%%%%%%%%%%%%%%%%%%%%%%



%\bibitem {BasteroGil:2009ec}M.~Bastero-Gil and A.~Berera,
%``Warm inflation model building,''
%Int.\ J.\ Mod.\ Phys.\ A \textbf{24}, 2207 (2009)



%\bibitem {Libro02}Arfken, G. "The Incomplete Gamma Function and Related
%Functions." Mathematical Methods for Physicists, 3rd ed. Orlando, FL: Academic
%Press, 1985.

%\bibitem {Bere2}A. Taylor and A. Berera, Phys. Rev. D \textbf{69}, 083517 (2000).

%\bibitem {fNL}S.~Gupta, A.~Berera, A.~F.~Heavens and S.~Matarrese,
%``Non-Gaussian signatures in the cosmic background radiation from warm
%inflation,''
%Phys.\ Rev.\ D \textbf{66}, 043510 (2002)

%\bibitem {fNL02}I.~G.~Moss and C.~Xiong,
%``Non-gaussianity in fluctuations from warm inflation,''
%JCAP \textbf{0704}, 007 (2007).


%\bibitem {17}M. C. Bento, O. Bertolami and A. Sen, Phys. Rev. D 66, 043507
%(2002).








%%%%%%%%%%%%%%%%%%%%Chaplygin








%\bibitem{14} M. C. Bento, O. Bertolami and A. Sen, Phys. Rev. D 66, 043507
%(2002).





%\bibitem{Bert}O. Bertolami and V. Duvvuri, Phys. Lett. B 640, 121 (2006).








%%%%%%%%%%%%%%%%intermediate

















\end{thebibliography}
\end{document}